\documentclass[aps,floatfix,superscriptaddress]{revtex4}
\usepackage{amsmath,amssymb,amsfonts,graphics,graphicx,dcolumn,bm,enumerate}
\usepackage{comment,natbib}
\usepackage[ddmmyyyy,hhmmss]{datetime}
\usepackage{multirow,color}

\newcommand{\cmp}
{\affiliation{Condensed Matter Physics Division, 
Saha Institute of Nuclear Physics, 1/AF Bidhannagar, Kolkata 700064, India.}}
\newcommand{\isi}
{\affiliation{Economic Research Unit, Indian Statistical Institute, 203 B. T. Road, Kolkata 700108, India.}}
\newcommand{\hok}
{\affiliation{Graduate School of Information Science \& Technology,  Hokkaido University, N14-W9, Kita-ku, Sapporo 060-0814, Japan.}}

\begin{document}
\title{Measuring social inequality with quantitative methodology:  analytical estimates and empirical data analysis by Gini and $k$ indices}

\author{Jun-ichi Inoue}
\email[Email: ]{jinoue@cb4.so-net.ne.jp, j_inoue@complex.ist.hokudai.ac.jp}
\hok
\author{Asim Ghosh}
\email[Email: ]{asim.ghosh@saha.ac.in}
\cmp
\author{Arnab Chatterjee}%
\email[Email: ]{arnabchat@gmail.com}
\cmp
\author{Bikas K. Chakrabarti}%
\email[Email: ]{bikask.chakrabarti@saha.ac.in}
\cmp \isi


\begin{abstract}
Social inequality manifested across different strata of human existence can be quantified in several ways.
Here we compute non-entropic measures of inequality such as Lorenz curve, Gini index and the recently introduced
$k$ index analytically from known distribution functions.
We characterize the distribution functions of different quantities such as votes, journal citations, city size, etc. with suitable fits,
compute their inequality measures and compare with the analytical results. 
A single analytic function is often not sufficient to fit the entire range of the 
probability distribution of the empirical data, and 
fit better to two distinct functions with a single crossover point.
Here we provide general formulas to calculate these inequality measures  
for the above cases.
We attempt to specify the crossover point by minimizing the gap between empirical and analytical evaluations of measures. 
Regarding the $k$ index as an `extra dimension', both the lower and upper bounds of the Gini index are obtained 
as a function of the $k$ index. This type of inequality relations among inequality indices might help us to check the validity of 
empirical and analytical evaluations of those indices. 
\end{abstract}

\maketitle
\section{Introduction}
Humans are social beings and our social interactions are often complex.
Social interactions in many forms produce spontaneous variations manifested as inequalities 
while at times these inequalities result out of continued complex interactions among the constituent human units.
The availability of a large body of empirical data for a variety of measures from human social interactions
has made it possible to uncover the patterns and investigate the reasons for socio-economic inequalities.
With tools of statistical physics as a core, researchers are incorporating the knowledge and 
techniques from various disciplines~\cite{lazer09} like statistics, applied mathematics,
information theory and computer science for a better
understanding of the nature and origin of socio-economic inequalities that shape the humankind.
Socio-economic inequality~\cite{arrow2000meritocracy,stiglitz2012price,neckerman2004social,goldthorpe2010analysing} 
is the existence of unequal opportunities and rewards for various social
positions or statuses within the society. It usually contains structured and recurrent
patterns of unequal distributions of goods, wealth, opportunities, and
even rewards and punishments, and mainly measured in terms of  
\textit{inequality of conditions},
and \textit{inequality of opportunities}.
\textit{Inequality of conditions} refers to the unequal distribution of income,
wealth and material goods. 
\textit{Inequality of opportunities} refers to the unequal distribution of `life
chances' across individuals. This is reflected in measures such as level of
education, health status, and treatment by the criminal justice system.
Socio-economic inequality is responsible for conflict, war, crisis, oppression, criminal activity,
political unrest and instability, and indirectly affects economic growth~\cite{hurst1995social}.
Traditionally, economic inequalities have been studied in the context of income and wealth~\cite{yakovenko2009colloquium,chakrabarti2013econophysics,aoyama2010econophysics}.
The study of inequality in society~\cite{Cho23052014,Chin23052014,Xie23052014}
is a topic of current focus and global interest and brings together
researchers from various disciplines -- economics, sociology, mathematics, statistics,
demography, geography, graph theory, computer science and even theoretical physics.

Socio-economic inequalities are quantified in various ways. The most popular measures are absolute,
in terms of indices, e.g., Gini~\cite{gini1921measurement}, Theil~\cite{theil1967economics}, Pietra~\cite{eliazar2010measuring} indices.
The alternative approach is a relative measure,
in terms of probability distributions of various quantities, but the most of the above mentioned indices
can be computed from the distributions. Most quantities often 
display broad distributions, usually lognormals, power-laws or their combinations.
For example, the distribution of income is usually an exponential followed by a power 
law~\cite{druagulescu2001exponential} (see Ref.\cite{chakrabarti2013econophysics} for other examples).

\begin{figure}[t]
\begin{center}
\includegraphics[width=9.0cm]{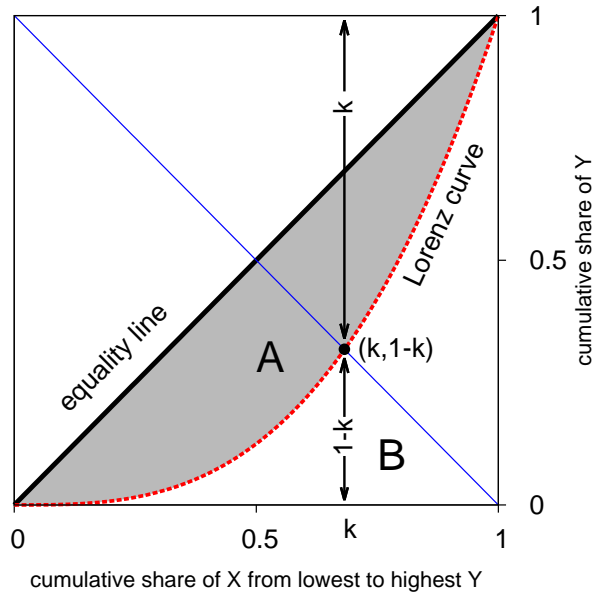}
\end{center}
\caption{
Schematic representation of 
Lorenz curve, Gini index $g$ and $k$ index.
The dashed red line stands for the Lorenz curve and the black solid line represents perfect equality.
The area enclosed by the equality line and the Lorenz curve is $A$ and that below the Lorenz curve is $B$.
The Gini index is given by $g=A/(A+B)$.
The $k$ index is given by the abscissa of the intersection point of the Lorenz curve and $Y=1-X$.
}
\label{fig:fg_pic_Lorenz}
\end{figure}

The Lorenz curve~\cite{Lorenz} is function which represents the cumulative proportion $X$ of 
ordered individuals (from lowest to highest) 
in terms of the cumulative proportion of their size $Y$.
Here, $X$ can represent income or wealth, citation, votes, city population etc. Table~\ref{tab:xy} shows the typical
examples of $X$ and the corresponding $Y$.
The Gini index ($g$) is defined as the ratio between the area enclosed between
the Lorenz curve and the equality line, and the area below the equality line.
If the area between 
(i) the Lorenz curve and the equality line is $A$, and 
(ii) that below the Lorenz curve is $B$  (See Fig.~\ref{fig:fg_pic_Lorenz}),
the Gini index is given by $g=A/(A+B)$.
It is an useful measure for quantifying socio-economic inequalities. 
Besides these well-established measures, 
Ghosh et al.~\cite{ghosh2014inequality}
recently introduced a different measurement 
called `$k$ index' (`$k$' stands for the extreme nature of social inequalities in Kolkata) 
defined as the fraction $k$ such that the cumulative income or citations of $(1-k)$ fraction 
of people or papers are held by fraction $k$ of the people or publications respectively.
\begin{table}[t]
\caption{Table showing examples of what $X$ and $Y$ can represent.}
\label{tab:xy}
\begin{tabular}{ |c|c| }
\hline
$X$ & $Y$ \\
\hline
people & income, wealth \\
article/paper & citation \\
institution/university & citations\\
institution/university & funding\\
candidate & vote \\
city & population \\
student & marks \\
company & employee \\
\hline
\end{tabular}
\end{table}

When the probability distribution is described using an appropriate parametric function, 
one can derive these inequality measures as a function of those parameters analytically. 
In fact, several empirical evidence have been reported to show 
that the distributions can be put into a finite number of types. 
Most of them turn out to be a of mixture of two distinct parametric 
distributions with a single crossover point.  

In this paper, we have characterized empirical data
and the fitting forms have been treated analytically for comparison.
We show in this paper that  
the distributions of population in socio-economic sciences can be put into 
several categories.  
We specify each of the distributions by appropriate parameters. 
We present  the general form of the inequality measures, 
namely, Lorenz curve, Gini index $g$ and $k$ index for a class of distributions 
which can be expressed as a mixture of two distributions with a single crossover point. 
We check the values obtained from empirical calculations with 
those from analytical expressions. 
Especially, by minimizing 
the empirical and analytical values of 
the inequality measures, 
one can find an estimate of the crossover point 
which is usually determined by eye estimates. 
As a use of $k$ index, 
both the  lower and upper bounds of the Gini index are obtained 
as a function of $k$ index by 
considering the $k$ index as an `extra dimension'. 
This type of inequality relation among the inequality indices might help us to check the validity of 
empirical and analytical estimates of these indices.

This paper is organized as follows. 
In Sec.~\ref{sec:Generic}, we introduce the basics and 
some generic properties of our 
measures -- Lorenz curve, Gini and $k$ indices. 
In Sec.~\ref{sec:general}, 
we provide the general formulas of the inequality measures  
for the empirically observed distributions. 
In Sec.~\ref{sec:inequality}, 
regarding the $k$ index as an `extra dimension', both  the lower and upper bounds of the Gini index are obtained 
as a function of $k$ index. 
In Sec.~\ref{sec:results}, we report our results. 
Here, we provide some empirical findings. 
Out of the data we considered, we found  six categories of  distributions.
We observed that 
most of the data can be described by a mixture of two distinct distributions 
with a crossover point. 
Here we give numerical evaluations of our measures.  
Then, we compare the empirical and analytical evaluations of 
inequality measures. 
Minimizing the gap between two results obtained by 
different ways, we infer the best possible crossover point 
for a given data set.  
We conclude with a summary and discussions. 
\section{Basics and generic properties of inequality measures}
\label{sec:Generic}
In this section, we introduce the measures to quantify 
the degree of social inequality, namely, 
Lorenz curve, Gini index and $k$ index. 
Then, the generic properties are explained. 
\subsection{Lorenz curve}
The Lorenz curve is given as a 
relationship between 
the cumulative distribution and 
the cumulative first moment of $P(m)$. 
Namely, for the mean and the normalized first moment  of $P(m)$, 
\begin{equation}
 X (r ) =  \int_{m_{0}}^{r}P(m)dm,\,\,\,
 Y (r ) =  
 \frac{\int_{m_{0}}^{r}mP(m)dm}
 {\int_{m_{0}}^{\infty} mP(m)dm}.
 \label{eq:def_Lorenz}
 \end{equation}
 The  Lorenz curve is given as 
 a set of $\left(X(r ),Y(r )\right)$, 
 where we assume that 
 the $P(m)$ is defined in $[m_{0},\infty)$. 
 In Fig.~\ref{fig:fg_pic_Lorenz}, we show the typical 
behavior of Lorenz curve by dashed line. 

 The intuitive meaning of the Lorenz curve is as follows: 
the cumulative proportion $X$ of ordered (from lowest to highest) individuals hold 
the cumulative proportion  $Y$ of wealth.
For sake of simplicity, we will use `individuals' for attributing $X$ and `wealth' for attributing $Y$,
for the simple reason that Lorenz curve, Gini index etc. were historically introduced in the context of income/wealth,
but in principle the attributes $X$ and $Y$ can be any of the combinations mentioned in Table.~\ref{tab:xy}.
 Hence,  when all individuals take the same amount 
of wealth, say $m_{*}$, we have 
\begin{equation}
P(m)=\delta (m-m_{*}), \quad \quad m_{0} <m_{*} <\infty,
\label{eq:Pm_equal}
\end{equation}
and one obtains 
\begin{eqnarray}
X (r ) & = & \int_{m_{0}}^{r} \delta (m-m_{*})dm =\Theta (r-m_{*}), \\
Y(r ) & = & \frac{\int_{m_{0}}^{r} m\delta (m-m_{*})dm}
{\int_{m_{0}}^{\infty}m\delta(r-m_{*})dm}
=\frac{m_{*}\Theta (r-m_{*})}{m_{*}}=X (r ). 
\end{eqnarray}
where 
$\Theta (x)$ is a unit step function defined by 
\begin{equation}
\Theta (x) = 
\left\{
\begin{array}{cc}
1, & \quad \quad x \geq 1, \\
0, & \quad \quad x < 1.
\end{array}
\right.
\end{equation}
Thus we have $ Y=X$ 
 as the `perfect equality line' (see thick line in Fig.~\ref{fig:fg_pic_Lorenz}). This means that  
 $X$ fraction of people takes $X$ fraction of total wealth in society. 
 
On the other hand, 
when the total wealth in the society consisting of $N$ persons is concentrated to a few persons, 
namely, 
\begin{equation}
P(m)=(1-\varepsilon) \delta_{m,0}+\varepsilon \delta_{m,1},
\label{eq:Pm_unequal}
\end{equation}
where $\varepsilon \sim \mathcal{O}(1/N)$ and 
we assume that the total amount of wealth is normalized as $1$, 
we obtain $X(r )=1-\varepsilon +\varepsilon \delta_{r,1}$ and 
$Y(r )=\delta_{r,1}$. 
Hence, $Y=1$ iff  $X=r=1$ and $Y=0$ otherwise, 
and 
the Lorenz curve is given as 
`perfect inequality line' by $Y=\delta_{X,1}$
where $\delta_{x,y}$ is a Kronecker's delta (see Fig.~\ref{fig:fg_pic_Lorenz}). 
\subsection{Gini index}
For a given Lorenz curve, the Gini index is evaluated as 
twice of area between the curve $(X(r ),Y(r ))$ and perfect equality line $Y=X$. 
The area is shown in the shaded part (named `{\sffamily A}') in Fig.~\ref{fig:fg_pic_Lorenz}. 
Namely, it reads 
\begin{equation}
g= 2 \int_{0}^{1}(X-Y)dX = 
2 \int_{r_{0}}^{\infty}(X(r )-Y(r ))\frac{dX}{dr}dr,
\label{eq:def_G}
\end{equation}
where we should keep in mind that $X^{-1}(0)=r_{0}, X^{-1}(1)=\infty$ should hold. 
In a graphical way, the Gini index is given as a ratio of two areas 
(`{\sffamily A}' and `{\sffamily B}') by $g={\rm A}/({\rm A}+{\rm B})$. 
From the definition, the Gini index $g$ is zero for perfect equality and unity for 
perfect inequality. 
It should be noted that the Gini index may be evaluated analytically when 
the distribution of population is obtained in a parametric way. 

In fact, 
in the references \cite{sazuka2007fluctuations,sazuka2009distribution}, 
in the context of analysis of waiting time (duration) of time series, 
the Gini index was analytically calculated for a parametric distribution. 
In \cite{sazuka2007fluctuations,sazuka2009distribution}, 
the so-called Weibull distribution was selected 
to quantify the inequality of duration $t$ 
of financial time series. 
The Weibull distribution is described by 
\begin{equation}
P_{\mu,\eta}(t)=\frac{\mu t^{\mu-1}}{\eta}
{\exp}
\left(
-\frac{t^{\mu}}{\eta}
\right).
\label{eq:Weibull}
\end{equation}
The resulting Gini index for the Weibull distribution is given as 
\begin{equation}
g=1-
\left(
\frac{1}{2}
\right)^{\mu}.
\label{eq:Gini_Weibull}
\end{equation}
The Weibull distribution $P_{\mu,\eta}(t)$ is identical to 
exponential distribution $\sim {\rm e}^{-t/\eta}$ for $\mu=1$, 
which means that the point process 
specified by exponentially distributed duration $t$ between events follows a Poisson process. 
Hence, we are confirmed that 
$g=1/2$ for $\mu=1$, and 
the deviation of the Gini index from $1/2$ for arbitrary 
process shows to what extent the resulting time series is different  from randomly generated events. 
\subsection{$k$ index}
The $k$ index which was recently introduced is defined as 
the value of $X$-axis for 
the intersection between 
the Lorenz curve and a straight line $Y=1-X$. 
Namely, 
for the solution of equation
\begin{equation}
X(r )+Y(r )=1,
\end{equation}
say $r_{*}=Z^{-1}(1), Z(r )\equiv X(r ) + Y(r )$, 
the $k$ index is given by 
\begin{equation}
k =X(r_{*}).
\end{equation}
From the definition, the $k$ index denotes 
the situation in which 
$k$ fraction of people shares totally $(1-k)$ fraction of the wealth. 
Obviously, the $k$ index takes $1/2$ for perfectly equal society, 
whereas it takes $1$ for perfectly unequal society. 


The $k$ index 
is obviously 
easier to estimate by eyes 
in comparison with the Gini index (shaded area {\sffamily A} in Fig.~\ref{fig:fg_pic_Lorenz}). 
We will also discuss 
another use of the $k$ index 
by regarding the $k$ as an extra dimension in Sec.~\ref{sec:inequality}. 

Besides $g$ and $k$ indices,
Pietra's $p$ index \cite{eliazar2010measuring} and median index or $m$ index 
\cite{eliazar2014socialinequality} has been used as inequality measures derived from the Lorenz curve. 
The $p$ index is defined
as the maximal vertical distance between the Lorenz curve and the line of perfect equality $Y=X$ (in Fig.~\ref{fig:fg_pic_Lorenz}), 
whereas $m$ index is given by $2m-1$ 
for the solution of $Y(m)=1/2$, where we assumed that the Lorenz curve is 
given as $(X(r ), Y(r ))$ using a parameter $r$. 
It might be important for us to discuss these two indices, 
however, in this paper we limit ourselves to 
$g$ and $k$ indices. 
\section{General formula}
\label{sec:general}
In this section, we describe the general formula 
for the Lorenz curve, Gini index and $k$ index for 
the distribution $P(m)$. 
In this paper, we calculate the $g$ and $k$ indices using different theoretical distribution
functions, and sometimes combinations of two of them. It is very common to find that the probability
distributions of many quantities (wealth, income, votes, citations etc.)
fit to more than one theoretical function depending on the range: 
\begin{equation}
P(m) = F_{1}(m)\theta(m,m_{\times})+
F_{2}(m)\Theta (m-m_{\times}),
\label{eq:Pm}
\end{equation}
$\theta (m,m_{\times}) \equiv \Theta (m)-\Theta (m-m_{\times})$, where  
$m_{\times}$ is the crossover point. 

The functions $F_1(m)$ and $F_2(m)$ are suitably normalized
and computed for their continuity at $m_{\times}$. 
We will compute the functional fits to the empirical distributions of several quantities,
compute the $g$ and $k$ indices, and compare with the theoretical values computed from
the fitting distributions.
\subsection{Lorenz curve}
Our main purpose here is to derive 
a general form of the Lorenz curve for 
the distribution having the form Eq.~(\ref{eq:Pm}). 
The resulting form of the Lorenz curve 
is given by 
\begin{equation}
Y = 
\left\{
\begin{array}{lc}
\frac{R_{1}[
Q_{1}^{-1}
(\{Q_{1}(m_{\times})+Q_{2}(m_{\times})\}X)
]}
{R_{1}(m_{\times})+R_{2}(m_{\times})}, & 
\quad \quad 
0 \leq X \leq 
\frac{Q_{1}(m_{\times})}
{Q_{1}(m_{\times})+Q_{2}(m_{\times})},
\\
1-
\frac{R_{2}[
Q_{2}^{-1}(\{Q_{1}(m_{\times})+Q_{2}(m_{\times})\}(1-X))]}
{R_{1}(m_{\times})+R_{2}(m_{\times})}, & 
\quad \quad 
\frac{Q_{1}(m_{\times})}
{Q_{1}(m_{\times})+Q_{2}(m_{\times})} \leq X \leq 1,
\end{array}
\right.
\label{eq:general_L}
\end{equation}
where we defined the cumulative `persons'  and `wealth'  of the distributions 
$F_{1}(m)$ and $F_{2}(m)$ as 
\begin{eqnarray}
Q_{1} (r ) & = & 
\int_{m_{0}}^{r}F_{1}(m)dm,\,\,\,
Q_{2}(r ) = 
\int_{r}^{\infty}
F_{2}(m) dm,
\label{eq:def_Q1Q2} \\
R_{1}(r ) & = & 
\int_{m_{0}}^{r}mF_{1}(m)dm,\,\,\,
R_{2}(r ) = 
\int_{r}^{\infty}mF_{2}(m)dm.
\label{eq:def_R1R2} 
\end{eqnarray}
The derivation is given in Appendix \ref{app:ap_der_general}. 

It should be noticed that 
when the mean and the first moment of $F_{1}(m)$ and $F_{2}(m)$ are identical in such a way as  
$Q_{1} (r ) =R_{1} (r )$ and $Q_{2} (r )=R_{2}(r )$; 
for instance, $F_{1}(m)$ and $F_{2}(m)$ are both $P(m)=\delta (m-m_{*}),\,\,0<m_{*}<\infty$, we have 
\begin{equation}
Y = \frac{R_{1}[
Q_{1}^{-1}
(\{Q_{1}(m_{\times})+Q_{2}(m_{\times})\}X)
]}
{R_{1}(m_{\times})+R_{2}(m_{\times})} 
= 1-
\frac{R_{2}[
Q_{2}^{-1}(\{Q_{1}(m_{\times})+Q_{2}(m_{\times})\}(1-X))]}
{R_{1}(m_{\times})+R_{2}(m_{\times})}  =X , 
\end{equation}
which is nothing but the Lorenz curve for perfect equal society. 

The above argument is very general and independent of specific choice of the 
distributions. However, it is important for us to check the validity of the above general form Eq.~(\ref{eq:general_L}) for 
well-known limiting cases 
without crossover, namely, 
$m_{\times} \gg 1$ or $m_{\times}=m_{0}$. 
\subsubsection{Uniform distribution}
For this purpose, we first examine 
a single uniform distribution 
$F_{1}(m)=1/a$, $m_{\times}=a \gg1, m_{0}=0$. 
For this case, we find 
$Q_{1}(m_{\times})=1, Q_{1}(m_{0})=Q_{2}(\infty)=Q_{2}(m_{\times})=0$ and 
\begin{equation}
Q_{1} (r ) =\frac{r}{a},\,\,
R_{1} (r )= \frac{r^{2}}{2a}.
\label{eq:QRinv}
\end{equation}
Those lead to 
$Q_{1}^{-1}[\{Q_{1}(m_{\times})+Q_{2}(m_{\times})\}X]=aX, 
R_{1}(
Q_{1}^{-1}[\{Q_{1}(m_{\times})+Q_{2}(m_{\times})\}X])=
R_{1}(aX)=aX^{2}/2$ and 
$R_{1}(m_{\times})=a/2 \gg 1, R_{2}(m_{\times})=0$. 
Hence, we obtain the Lorenz curve from the first branch of Eq.~(\ref{eq:general_L}) as 
\begin{equation}
Y = X^{2}. 
\label{eq:Lorenz_uniform}
\end{equation}
\subsubsection{Power law distribution}
We next consider 
the case of $m_{\times} \gg 1$ and 
$F_{1}(m)=(\alpha-1)m^{-\alpha}, m_{0}=1$, 
namely, for a single power law distribution. 
Then, 
the first branch in Eq.~(\ref{eq:general_L}) survives and 
we have 
$Q_{1}(m_{\times})=1, 
R_{1}(m_{\times})=(\alpha-1)/(\alpha-2),
Q_{2}(m_{\times})=R_{2}(m_{\times})=0$, 
and $Q_{1} (r )=1-r^{1-\alpha}=(Q_{1}(m_{\times})+Q_{2}(m_{\times}))X=X$, 
namely, 
\begin{equation}
Q_{1}^{-1}(\{Q_{1}(m_{\times})+Q_{2}(m_{\times})\}X) = (1-X)^{\frac{1}{1-\alpha}}.
\label{eq:Q1inv}
\end{equation}
Therefore, 
using $R_{1}(r ) = (\alpha-1) (1-r^{2-\alpha})/(\alpha-2)$, 
we have
\begin{equation}
R_{1}[Q_{1}^{-1}(\{Q_{1}(m_{\times})+Q_{2}(m_{\times})\}X)] = 
\left(
\frac{\alpha-1}{\alpha-2}
\right)
\{
1-(1-X)^{\frac{2-\alpha}{1-\alpha}}
\}.
\end{equation}
Inserting these staffs into 
the first branch of Eq.~(\ref{eq:general_L}), we finally obtain
\begin{equation}
Y=1-(1-X)^{\frac{2-\alpha}{1-\alpha}}. 
\end{equation}
\subsubsection{Lognormal distribution}
We next consider the case of $m_{\times} \gg 1, m_{0}=0$ and 
$F_{1}(m)$ is a lognormal distribution given by 
\begin{equation}
F_{1}(m) = 
\frac{1}{\sqrt{2\pi}\sigma m}
{\exp}
\left[
-\frac{(\log m -\mu)^{2}}{2\sigma^{2}}
\right]. 
\label{eq:a_lognormal}
\end{equation}
For this case, the first branch in Eq.~(\ref{eq:general_L}) is 
selected and 
$Q_{1}(m_{\times})=1$, $R_{1}(m_{\times})={\rm e}^{\mu + \sigma^{2}/2}$, 
$Q_{2}(m_{\times})=R_{2}(m_{\times})=0$. 
We also have 
\begin{equation}
Q_{1} (r ) = H
\left(
\frac{\mu -\log r}{\sigma}
\right),\,\,\,
R_{1}(r ) = 
{\rm e}^{\mu + 
\frac{\sigma^{2}}{2}}
H 
\left(
\frac{\mu + \sigma^{2}-\log r}{\sigma}
\right).
\end{equation}
where we defined
\begin{equation}
H(x)=\int_{x}^{\infty}Dz,\,\,\,
Dz \equiv \frac{dz}{\sqrt{2\pi}}{\rm e}^{-z^{2}/2}. 
\end{equation}
This reads 
\begin{eqnarray}
Q_{1}^{-1}(\{Q_{1}(m_{\times})+Q_{2}(m_{\times})\}X) & = & {\exp}
[\mu-\sigma H^{-1}(X)] ,
\label{eq:Q1inv2}\\
R_{1}[Q_{1}^{-1}(\{Q_{1}(m_{\times})+Q_{2}(m_{\times})\}X)] & = & 
{\rm e}^{\mu + \frac{\sigma^{2}}{2}}
H(\sigma + H^{-1}(X)). 
\end{eqnarray}
Substituting these all staffs to the first branch of Eq.~(\ref{eq:general_L}), 
finally we obtain 
\begin{equation}
Y = H(\sigma + H^{-1}(X)).
\end{equation}
This reads 
$H^{-1}(Y)=\sigma + H^{-1}(X)$. 
Hence, we recover the result for perfect equality $Y=X$ in 
the limit of $\sigma \to 0$. 
\subsection{Gini index}
For the general distribution with a crossover Eq.~(\ref{eq:Pm}), 
we can derive the general form of the Gini index as follows. 
\begin{equation}
 g  =  
 \frac{Q_{1}(m_{\times})^{2}-
 Q_{1}(m_{0})^{2}+
 Q_{2}(\infty)^{2}-Q_{2}(m_{\times})^{2}}
 {
 \{
 Q_{1}(m_{\times})+Q_{2}(m_{\times})
 \}^{2}
 } - 
 \frac{2 
 (S_{1}(m_{0},m_{\times})+
 T_{2}(m_{\times}))
 }
 {
  \{
 Q_{1}(m_{\times})+Q_{2}(m_{\times})
 \}
  \{
 R_{1}(m_{\times})+R_{2}(m_{\times})
 \}
 },
 \label{eq:general_G}
 \end{equation}
where we defined 
\begin{equation}
S_{1}(m_{0},m_{\times})=\int_{m_{0}}^{m_{\times}}
 R_{1}(r ) 
 \frac{dQ_{1} (r )}{dr}
 dr; \quad 
 T_{2}(m_{0})=
\int_{m_{\times}}^{\infty}
 R_{2}(r ) 
 \frac{d Q_{2} (r )}{dr}
 dr. 
 \label{eq:def_S1T2}
\end{equation}
\mbox{}
We should keep in mind that 
we replace $\infty$ 
in the upper bound of integral in $T_{2}(m_{0})$ and 
$Q_{2}(\infty)$ 
by $M <\infty$ 
when the distribution function 
$F_{2}(m)$ possesses a cut-off $M$. 
The detail of the derivation is explained in Appendix \ref{app:ap_der_general}. 

To check the validity of the general form Eq.~(\ref{eq:general_G}), 
we examine the case of a single power law distribution and 
a single lognormal distribution as we did for checking the Lorenz curve. 
\subsubsection{Uniform distribution}
We first examine a single uniform distribution 
$F_{1}(m)=1/a, m_{\times}=a \gg 1$, $m_{0}=0$. 
Taking into account the result 
$S_{1}(m_{0},m_{\times})= 2a/3$, $T_{2}(m_{\times})=0$, we obtain 
from the general form Eq.~(\ref{eq:general_G}) as 
\begin{equation}
g = \frac{1}{3}. 
\end{equation}
\subsubsection{Power law distribution}
For a power law distribution, 
we should set $m_{\times} \gg 1, m_{0}=1$ and we have 
 $S_{1}(m_{0},m_{\times})=(\alpha-2)/(2\alpha-3)$, $T_{2}(m_{\times})=0$
 Taking into account the result and 
 $Q_{1}(m_{0})=Q_{2}(\infty)=0$, we have 
 \begin{equation}
 g = \frac{1}{2\alpha-3}. 
 \end{equation}
 Therefore, 
 the $g$ for $\alpha=3$ is identical to 
 the result of uniform distribution $g=1/3$.  
 \subsubsection{Lognormal distribution}
 We next consider the case of a single lognormal distribution Eq.~(\ref{eq:a_lognormal}). 
 Here we notice 
 \begin{equation}
 S_{1}(m_{0},m_{\times})=
 {\rm e}^{\mu + \frac{\sigma^{2}}{2}}
 \int_{-\infty}^{\infty}Dz H(z+\sigma), \,\,\,
 T_{2}(m_{\times})=0
 \end{equation}
 and $Q_{1}(m_{0})=Q_{2}(\infty)=0$. Thus, we  obtain 
 \begin{equation}
 g = 1-2\int_{-\infty}^{\infty}
 Dz H(z+\sigma). 
 \end{equation}
 Using the fact  
 $\int_{-\infty}^{\infty}Dx H(x)=\{H(-\infty)\}^{2}/2=1/2$, the above expression is rewritten 
 in terms of the integral of difference between two complementary error functions as 
 \begin{equation}
 g = 
 2\int_{-\infty}^{\infty}
 Dx\{
 H(x)-H(x+\sigma)
 \}, 
 \end{equation}
and one can confirm that we recover the result for perfect equality case 
in the limit of $\sigma \to 0$ as $g=0$. 
On the other hand, in the limit of $\sigma \to \infty$, 
the mode of the lognormal behaves as ${\rm e}^{\mu -\sigma^{2}} \to 0$. 
Hence, the distribution might possess the form $P(m)=(1-\varepsilon)\delta_{m,0}+
\varepsilon \delta_{m,1}$, where $\varepsilon$ is a small fraction of people in 
the society, 
and here we suppose the maximum value of wealth is normalized as $1$ (see also Eq.~(\ref{eq:Pm_unequal})). 
Therefore, the Gini index should be identical to the value for perfectly unequal society, 
and actually we have
\begin{equation}
g = 
2\int_{-\infty}^{\infty}
Dx H(x) = 
2 \times \frac{1}{2}\{H(-\infty)\}^{2}=1. 
\end{equation}
\subsection{$k$ index}
The general form of the $k$ index for the distribution $P(m)$ (Eq.~(\ref{eq:Pm}) 
is given as a solution of the following equation: 
\begin{equation}
\left\{
\begin{array}{lc}
Q_{1}^{-1}(\{Q_{1}(m_{\times})+Q_{2}(m_{\times})\}k)=
R_{1}^{-1}(\{
R_{1}(m_{\times})+R_{2}(m_{\times})\}
(1-k)), & 
\quad \quad 
0 \leq k \leq 
\frac{Q_{1}(m_{\times})}
{Q_{1}(m_{\times})+Q_{2}(m_{\times})},
\\
Q_{2}^{-1}(\{Q_{1}(m_{\times})+Q_{2}(m_{\times})\}(1-k))=
R_{2}^{-1}(\{
R_{1}(m_{\times})+R_{2}(m_{\times})\}k), & 
\quad \quad 
\frac{Q_{1}(m_{\times})}
{Q_{1}(m_{\times})+Q_{2}(m_{\times})} <k \leq 1.
\end{array}
\right. 
\label{eq:general_k}
\end{equation}
The derivation is given in Appendix \ref{app:ap_der_general}. 

It should be noticed that 
for $Q_{1} (r ) =R_{1} (r )$ and $Q_{2} (r )=R_{2}(r )$; 
for instance, $F_{1}(m)$ and $F_{2}(m)$ are both $P(m)=\delta (m-m_{*}),\,\,0<m_{*}<\infty$, we have 
$k=1-k$. It reads $k = \frac{1}{2}$
which is the $k$ index for perfectly equal society. 
\subsubsection{Uniform distribution}
To check the validity, we next examine the case of 
$F_{1}(m)$ is a uniform distribution 
$F_{1}(m)=1/a$, $m_{\times}=a \gg 1$, $m_{0}=0$.  
From Eq.~(\ref{eq:QRinv}), we have 
$Q_{1}^{-1}(k)=ak$, 
$R_{1}^{-1}(a(1-k)/2)=a\sqrt{1-k}$. 
Thus, from the first branch of Eq.~(\ref{eq:general_k}), we obtain 
$k=\sqrt{1-k}$. That is 
\begin{equation}
k=\frac{-1+\sqrt{5}}{2} \sim 0.62. 
\label{eq:kindex_uniform}
\end{equation}
\subsubsection{Power law distribution}
We next consider the case in which  
$F_{1}(m)$ follows a power law distribution 
under the condition 
$m_{\times} \gg 1$, $m_{0}=1$. 
From Eq.~(\ref{eq:Q1inv}), we have 
$Q_{1}^{-1}(\{Q_{1}(m_{\times})+Q_{2}(m_{\times})\}k) = (1-k)^{\frac{1}{1-\alpha}}$. 
Using 
$R_{1}^{-1}(\{R_{1}(m_{\times})+R_{2}(m_{\times})\}(1-k))=k^{\frac{1}{2-\alpha}}$, 
the $k$ index is determined as a solution of 
\begin{equation}
k = (1-k)^{\frac{2-\alpha}{1-\alpha}}.
\end{equation}
In particular, we obtain for the choice of $\alpha=3$ as 
$k=\frac{-1+\sqrt{5}}{2} \sim 0.62$. 
We should notice that 
the value is exactly the same as that of uniform distribution. 
Hence, the $k$ index for a power low distribution with exponent $\alpha=3$ is 
identical to that of a uniform distribution as 
we saw it for $g$ index.  
\subsubsection{Lognormal distribution}
\begin{figure}[t]
\begin{center}
\includegraphics[width=8.9cm]{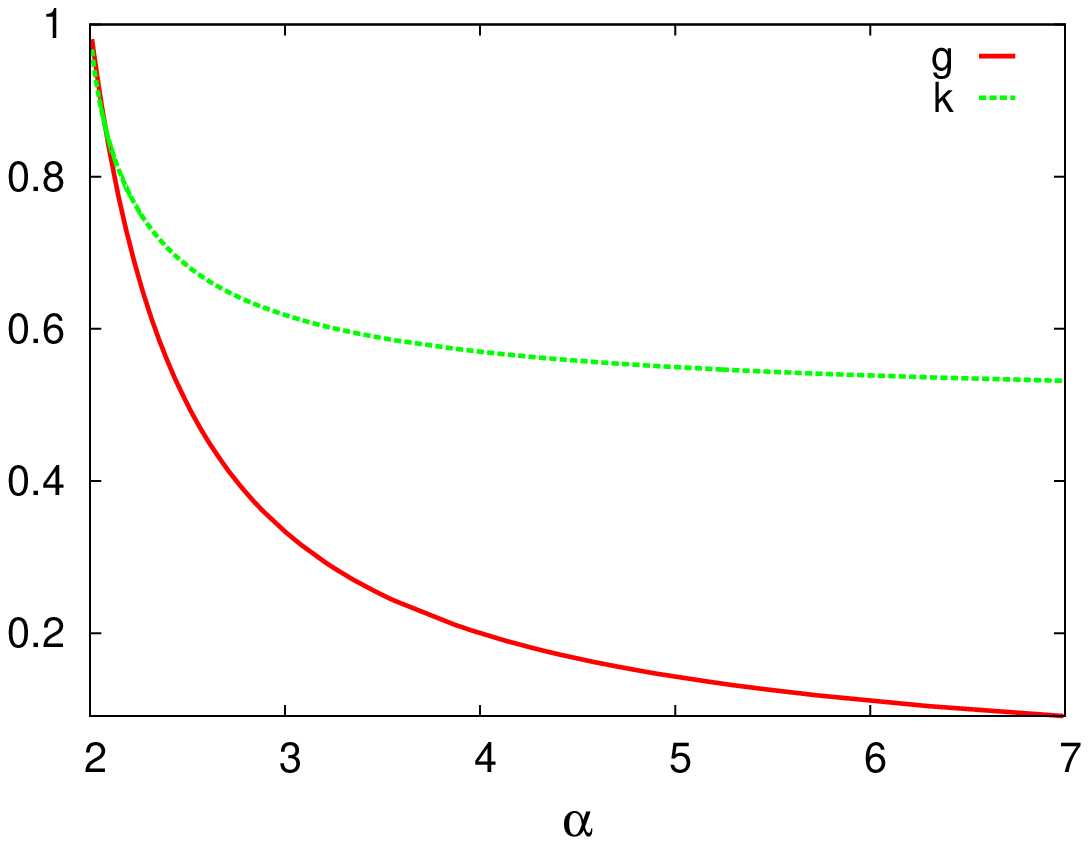}
\includegraphics[width=8.9cm]{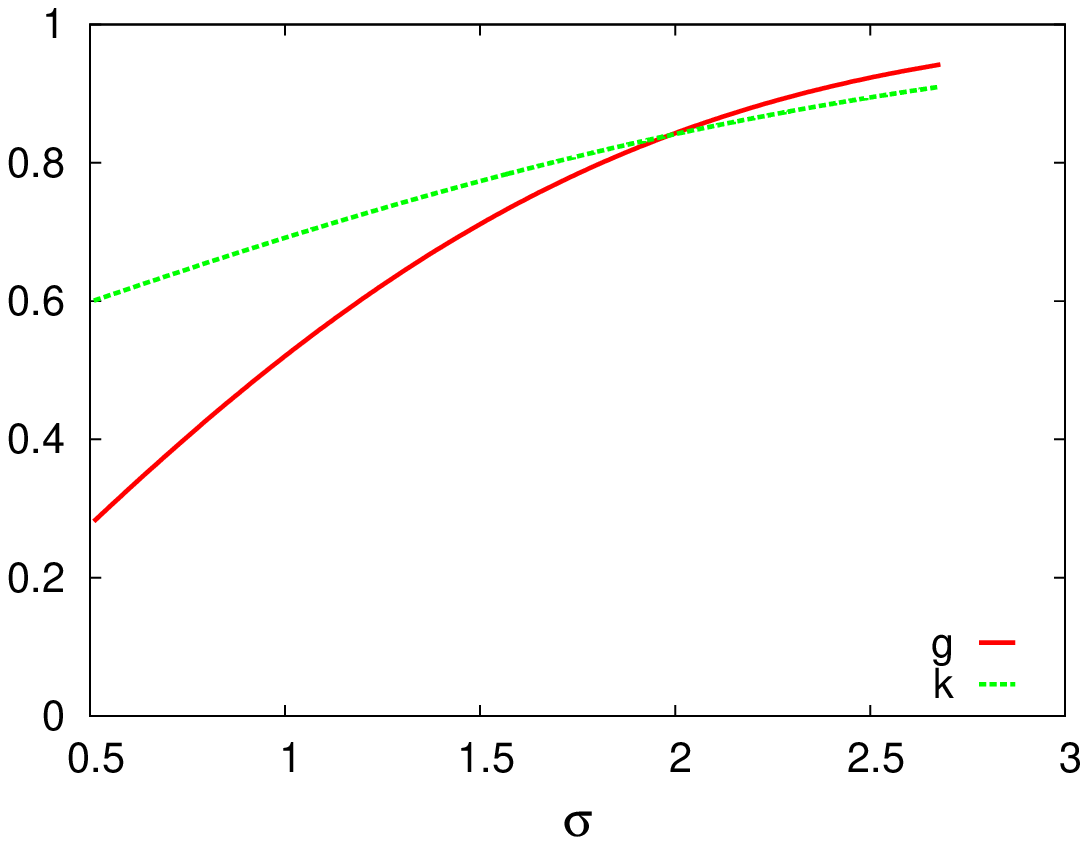}
\end{center}
\caption{
The $g$ and $k$ indexes 
for a single power law distribution (left) and 
a single lognormal distribution with $\mu=1$ (right: the case (a) in Fig.~\ref{fig:various1}) 
as a function of $\alpha$ and $\sigma$, 
respectively. 
For a single lognormal distribution, 
the $g$ goes to $0$ in the limit of $\sigma \to 0$,  
whereas the $k$ goes to $1/2$, each of which is 
a limit of `perfect equality'. 
On the other hand, 
both $g$ and $k$ go to $1$ in the limit of $\sigma \to \infty$. 
}
\label{fig:fg_kG_single}
\end{figure}

On the other hand, 
for a lognormal as $F_{1}(m)$ and 
$m_{\times}\gg 1, m_{0}=0$, we obtain 
from Eq.~(\ref{eq:Q1inv2}) as
$Q_{1}^{-1}(\{Q_{1}(m_{\times})+Q_{2}(m_{\times})\}k)  =  {\exp}
[\mu-\sigma H^{-1}(k)] $. 
Using the relation 
$R_{1}^{-1}(\{R_{1}(m_{\times})+R_{2}(m_{\times})\}(1-k))=
{\exp}[\mu + \sigma^{2}-\sigma H^{-1}(1-k)]$, we obtain 
the $k$ index as a solution of 
\begin{equation}
H^{-1}(1-k)-H^{-1}(k)=\sigma. 
\label{eq:k_lognormal}
\end{equation}
It should be noticed that 
from the definition of the lognormal distribution, 
we find $P(m)=\delta (m-{\rm e}^{\mu})$ in the limit of $\sigma \to 0$
(in this limit, the median, mean and mode of the lognormal take the same value ${\rm e}^{\mu}$). 
Namely, all person possesses the same wealth. In this limit, 
Eq.~(\ref{eq:k_lognormal}) leads to $H^{-1}(1-k)=H^{-1}(k)$, and 
this gives the $k$ index for 
perfectly equal society $k=1/2$. 
On the other hand, 
in the limit of $\sigma \to \infty$, 
the solution of Eq.~(\ref{eq:k_lognormal}) is 
$k=1$. 
We confirm these two limits in Fig.~\ref{fig:fg_kG_single} (right). 
\section{Upper and lower bounds for $g$ in terms of $k$}
\label{sec:inequality}
In the previous sections, we have introduced the Lorenz curve and 
$g$ and $k$ indices, and discussed several analytic forms and their properties.
These two indices $g$ and $k$ are both derived from the Lorenz curve and 
we might use some geometrical (graphical) interpretations for them.  
In particular, it might be very useful for us to derive the inequality for both 
inequality measures, 
that is,  `inequality of inequalities'. 

Here we derive the upper and lower bounds for the Gini index $g$ in terms of the $k$ index. 
The use of this lower and upper bounds
is one of the advantage of the  $k$ index.
For human eyes, it is very difficult to estimate the `area' 
surrounded by $Y=X$ and the Lorenz curve, that is, 
a half of Gini index $g/2$, whereas the $k$ index,
it is relatively easer for us to estimate the value by eyes because 
each axis $X$ and $Y$ is 
calibrated in $[0,1]$. 
In this sense, once we obtain both  the  bounds as a function of $k$,
we can easily estimate the $g$ from both  the  bounds through the $k$ index.
Additionally, if we can make the bound a tighter one,
the estimation will be closer to the exact value. 
The argument that follows is just an application of
basic Euclidean geometry.

\begin{figure}[ht]
\begin{center}
\includegraphics[width=9.0cm]{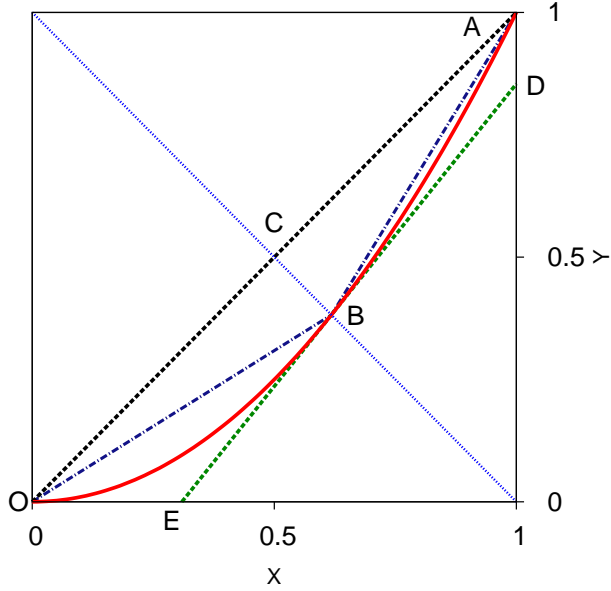}
\end{center}
\caption{
The case of uniform distribution. 
The Lorenz curve is simply given by $Y=X^{2}$ (see Eq.~(\ref{eq:Lorenz_uniform})). 
As the $k$ index is given by $k=(-1+\sqrt{5})/2$ (see Eq.~(\ref{eq:kindex_uniform})), 
the tangential line is obtained as 
$Y=(-1+\sqrt{5})(X-(\sqrt{5}-1)/2)+(3-\sqrt{5})/2$, 
which touches with the Lorenz curve at point B. 
We compare the area of the quadrilateral 
OADEO and the half of the Gini index $g$. 
Apparently the former is bigger than the latter, 
which gives the upper bound $\psi (k,\xi(k))$ of $g$. 
On the other hand, 
when we compare the area of triangle OAB and 
$g/2$, the former is smaller than the latter, which gives 
the lower bound $\phi (k)$ of $g$.  
}
\label{fig:fg_bound_uniform}
\end{figure}

Let us denote the origin $(0,0)$ as `O',  
$(1,1)$ as `A', $(k,1-k)$ as `B', 
and the intersection of perfect quality line 
$Y=X$ and $Y=1-X$, that is, $(1/2,1/2)$ as `C' 
 in Fig.~\ref{fig:fg_pic_Lorenz} (or Fig.~\ref{fig:fg_bound_uniform} for uniform distribution as a special case). 
Then, we compare the area of the triangle OAB and the shaded area which 
gives a half of the Gini index $g/2$. 
Obviously, as long as the Lorenz curve is convex, 
the area of the triangle OAB, $k-1/2$, is smaller than that of the shaded area $g/2$. 
Hence, we have 
\begin{equation}
g \geq 2k-1. 
\label{eq:inequality}
\end{equation}
The above equality is valid for  $k=1$ (perfect unequal) and $k=1/2$ (perfect equal). 

The convexity of the Lorenz curve is proved as follows. 
From the definition of the Lorenz curve Eq.~(\ref{eq:def_Lorenz}), we immediately have 
\begin{equation}
\frac{dY}{dX}=\frac{\frac{dY}{dr}}{\frac{dX}{dr}}=\frac{rP(r )}{Y_{0}P(r )}=\frac{r}{Y_{0}}, 
\end{equation}
and we conclude 
\begin{equation}
\frac{d^{2}Y}{dX^{2}}=
\frac{d}{dX}
\left(
\frac{dY}{dX}
\right)=
\frac{d}{dr}\left(
\frac{r}{Y_{0}} 
\right)  \cdot 
\frac{1}{\frac{dX}{dr}}=\frac{1}{Y_{0}P(r )} >  0, 
\end{equation}
where we defined $Y_{0} \equiv \int_{m_{0}}^{\infty}mP(m)dm$ as a positive constant. 
Therefore, the Lorenz curve is convex at any point of $X (r )$, and the inequality Eq.~(\ref{eq:inequality}) 
is actually satisfied for any set of inequality measures $g$ and $k$ for a given set of data sets or parameters which specify the 
probability density $P(m)$.

We next derive the upper bound of the Gini index by means of 
the $k$ index. 
To derive the bound, 
we consider the tangential line of the Lorenz curve at $(k,1-k)$ in Fig.~\ref{fig:fg_pic_Lorenz} 
(or Fig.~\ref{fig:fg_bound_uniform} for uniform distribution as a special case), 
that is, 
\begin{eqnarray}
Y & = & 
\xi (k)
(X-k) +1-k,\\
\xi (k) & \equiv &  
\frac{dY}{dX}{\Biggr |}_{X=k} =   
\frac{X^{-1}(k)}{Y_{0}}. 
\label{eq:def_xi}
\end{eqnarray}
Then, let us define the intersections 
of this tangential line and $X=1$, 
namely, 
$(1,(1-k)(1+\xi(k)))$ as `D'
and $Y=0$, $(k-(1/\xi (k))(1-k),0)$ as `E', respectively. 
Then, the area of a quadrilateral OADEO is larger than or equal to a half of the Gini index. 
Hence, using this fact, we can derive another inequality. 
The area is easily calculated and we have 
\begin{equation}
g \leq 
2k(2-k)-1 -
(1-k)^{2}
\left(
\xi(k)+ 
\frac{1}{\xi(k)}
\right). 
\label{eq:upper_bound}
\end{equation}
We should notice that the upper bound of the Gini index gives 
$1$ for $k=1$, whereas for $k=1/2$, we have 
\begin{equation}
g \leq 
\frac{1}{2} 
-
\frac{1}{4}
\left(
\xi (1/2) + 
\frac{1}{\xi (1/2)}
\right). 
\end{equation}
From the definition Eq.~(\ref{eq:def_xi}), 
$\xi (1/2)=dY/dX|_{X=k}=1$ because 
$k=1/2$ means  
the perfect equality line $Y=X$. 
Thus, we conclude $g \leq 0$ (which means  $g=0$ from the definition of $g$) for $k=1/2$. 
Therefore, the equality in Eq.~(\ref{eq:upper_bound}) 
should hold if and only if $k=1$ (perfect unequal) and $k=1/2$ (perfect equal). 

From the argument above, we finally obtain the following inequality  
\begin{equation}
\phi (k) \equiv 2k-1 \leq g \leq 
2k(2-k)-1 -
(1-k)^{2}
\left(
\xi(k)+ 
\frac{1}{\xi(k)}
\right) \equiv \psi (k,\xi(k)),
\label{eq:upperlower}
\end{equation}
where we should notice that 
the lower bound $\phi (k)$ is dependent on the detail of the distribution $P(m)$ 
through $k$ index itself, whereas the upper bound $ \psi (k, \xi (k))$ depends on the wealth distribution $P(m)$ through 
$k$ and the slope $\xi (k)$. 
\mbox{}

To check the validity of the inequality Eq.~(\ref{eq:upperlower}), 
we first consider a uniform distribution. 
The situation is shown in Fig.~\ref{fig:fg_bound_uniform}. 
As we already saw, the Lorenz curve is given by $Y=X^{2}$ (see Eq.~(\ref{eq:Lorenz_uniform}))  
which gives $\xi(k)=2k$. 
Thus, the upper bound is simply given by 
\begin{equation}
\psi(k,\xi(k))=
-2k^{3}+2k^{2}+2k
-1
-\frac{(1-k)^{2}}{2k}.
\end{equation}
Substituting the $k$ index for a uniform distribution 
$k=(-1+\sqrt{5})/2$ (see Eq.~(\ref{eq:kindex_uniform})) into 
the bounds in Eq.~(\ref{eq:upperlower}) 
$\phi (k)$ and $\psi(k,1/2)$, we obtain 
\begin{equation}
2k-1=\sqrt{5}-2=0.2360 < g < \psi(k,1/2)=\frac{12-5\sqrt{5}}{2}=0.4098.
\end{equation}
We should notice that 
the exact value of $g=1/3=0.3333$ for a uniform distribution is lying on the interval 
suggested by inequality Eq.~(\ref{eq:upperlower}).

We next check the bounds for a single exponential distribution: 
$P(m)=\beta \,{\rm e}^{-\beta m}$. 
It it is easy to derive the Lorenz curve and we obtain 
\begin{equation}
Y=X+(1-X)\log (1-X).
\end{equation}
Therefore, it is independent of the parameter $\beta$. 
For the Lorenz curve, the $k$ index is obtained as a solution of equation 
$2k=1-(1-k)\log (1-k)$, and it leads to $k=0.6822$. 
Hence, taking into account 
$\xi(k)=-\log (1-k)$, 
we obtain 
\begin{equation}
\psi (k,-\log (1-k))=-2k^{2}+4k-1 +
(1-k)^{2}
\left\{
\log (1-k) + 
\frac{1}{\log (1-k)}
\right\}
\end{equation}
and the inequality for $g$ in terms of $k$ is given by 
\begin{equation}
\phi (k) =2k-1=0.3644 < g < \psi (k, -\log (1-k))=0.5940. 
\end{equation}
Exact value of the Gini index for a single 
exponential distribution $\beta \, {\rm e}^{-\beta m}$ is evaluated as $g=1/2$, 
which is of course independent on $\beta$ and a special case of 
Weibull distribution Eqs.~(\ref{eq:Weibull}), (\ref{eq:Gini_Weibull}) with $\mu=1$. 
Hence, we are confirmed that the above inequality on $g$ actually works 
for a single exponential distribution. 
\begin{figure}[ht]
\begin{center}
\includegraphics[width=8.9cm]{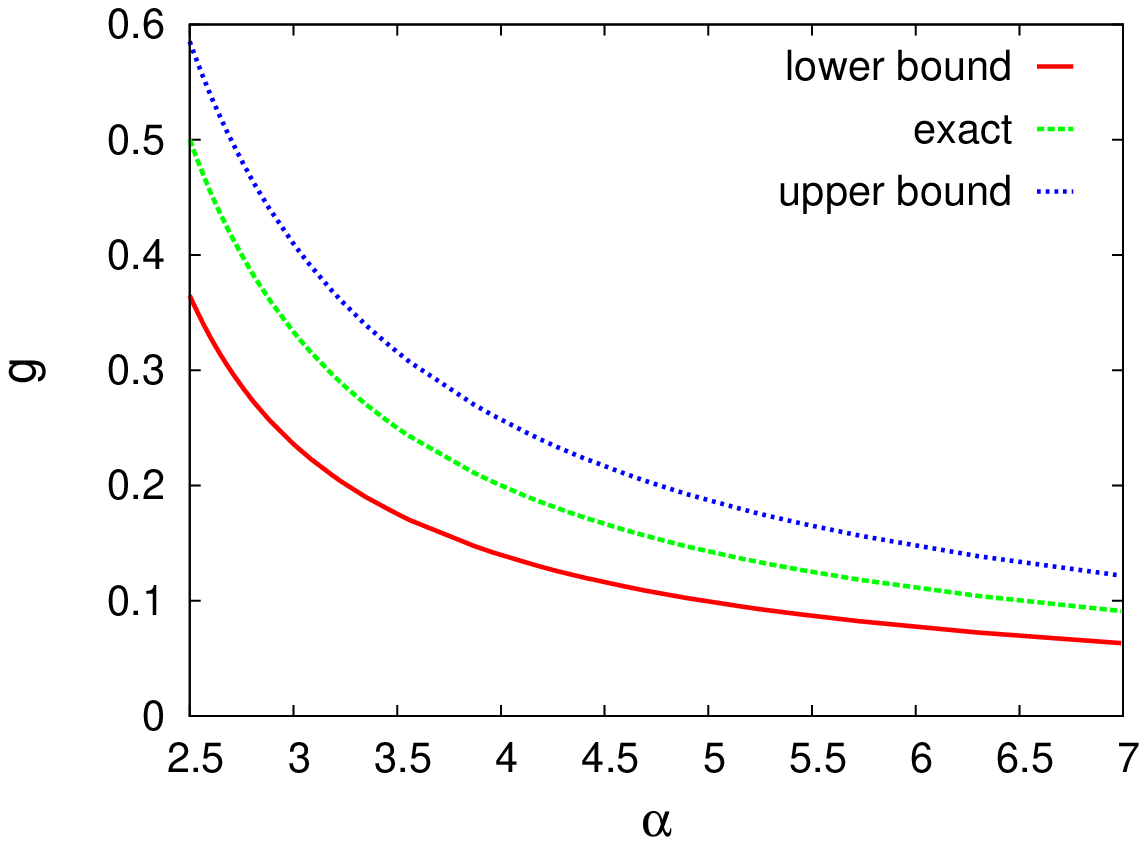}
\includegraphics[width=8.9cm]{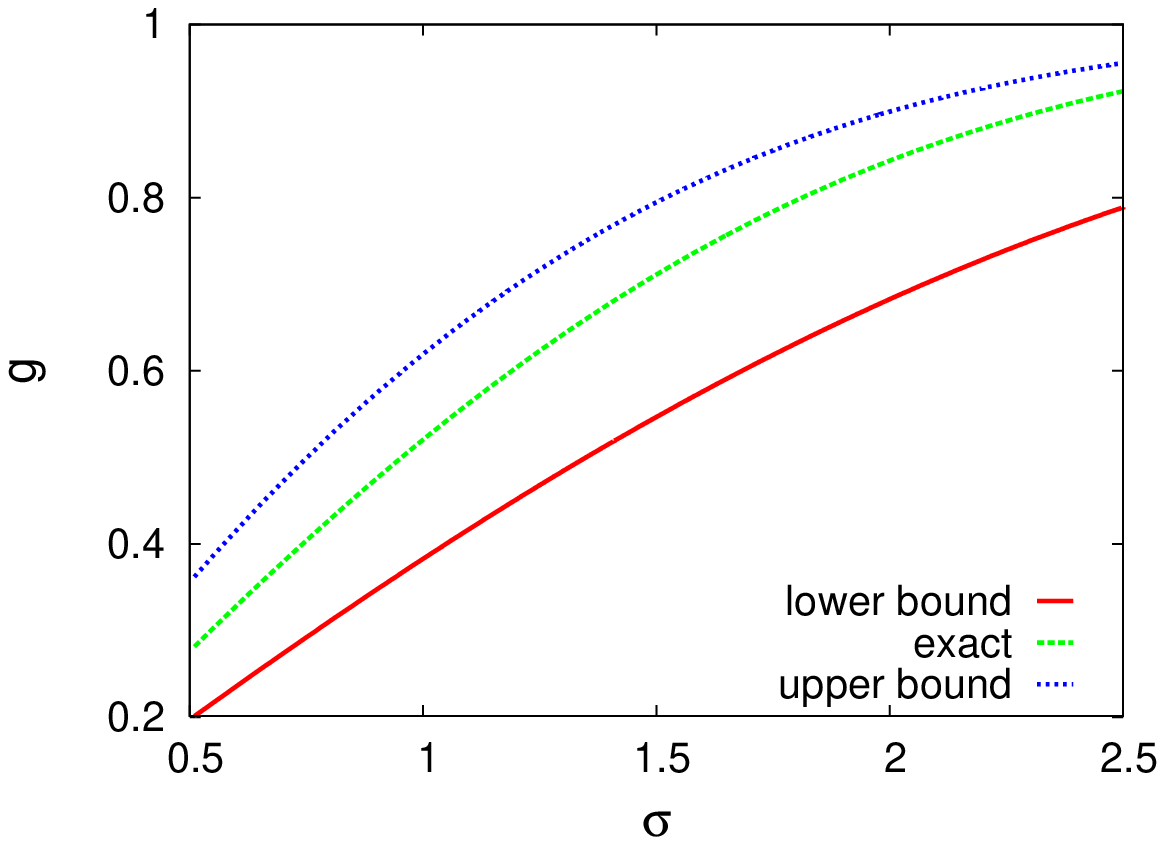}
\end{center}
\caption{
The upper and lower bounds of the Gini index 
for a single power law distribution (left) and 
a single lognormal distribution (right) 
with $\mu=1$  
as a function of parameters $\alpha$ and $\sigma$ respectively. 
The exact values of $g$ are the same as shown in Fig.~\ref{fig:fg_kG_single}. 
}
\label{fig:fg_g_bound}
\end{figure}

For the above two cases, 
both bounds $\phi (k), \psi(k,\xi(k))$ 
are independent of the parameters appearing in 
the wealth distribution $P(m)$. 
It should be stressed that 
the bounds are also evaluated 
even for parameter-dependent cases 
and $\phi (k), \psi(k,\xi(k))$ are obtained as a
function of parameters. 
As we shall show in the next section, 
most of the distributions of 
population in social science 
is categorized into several classes of distribution. 
Especially, 
those are described by a mixture of 
lognormal and power law distributions. 
Hence, it might be worthwhile for us to discuss 
the bounds for 
each of the distributions independently. 
We first consider the case of power law distribution: 
$P(m)(\alpha-1)\,m^{-\alpha}$. 
As we already obtain the Lorenz curve, 
we have the slope $\xi (k)=dY/dX|_{X=k}$ as 
\begin{equation}
\xi (k) = 
\left(
\frac{2-\alpha}{1-\alpha}
\right) (1-k)^{\frac{1}{1-\alpha}}. 
\end{equation}
Hence, we have 
\begin{equation}
\psi\left(
k, \left(
\frac{2-\alpha}{1-\alpha}
\right) (1-k)^{\frac{1}{1-\alpha}}
\right) = 
2k(2-k)-1-
(1-k)^{2}
\left\{
\left(
\frac{2-\alpha}{1-\alpha}
\right) (1-k)^{\frac{1}{1-\alpha}}
+
\left(
\frac{1-\alpha}{2-\alpha}
\right)
(1-k)^{-\frac{1}{1-\alpha}}
\right\}. 
\end{equation}
On the other hand, 
for a single lognormal distribution, 
we have 
\begin{equation}
\xi (k) = \frac{X^{-1}(k)}{{\rm e}^{\mu + \frac{\sigma^{2}}{2}}}. 
\end{equation}
Therefore, we have the upper bound as a function of $k$ as 
\begin{equation}
\psi \left(
k,
\frac{X^{-1}(k)}{{\rm e}^{\mu + \frac{\sigma^{2}}{2}}}
\right) = 
2k(2-k)-1-
(1-k)^{2} 
\left\{
\frac{{\rm e}^{\mu + \frac{\sigma^{2}}{2}}}{X^{-1}(k)}
+ 
\frac{X^{-1}(k)}
{{\rm e}^{\mu + \frac{\sigma^{2}}{2}}}
\right\}. 
\end{equation}
In Fig.~\ref{fig:fg_g_bound},  
we draw the upper and lower bounds for 
a single power law distribution (left) and 
a single lognormal distribution (right) 
with $\mu=1$  
as a function of parameters $\alpha$ and $\sigma$, respectively. 
From this figure, we find that 
the inequality Eq.~(\ref{eq:upperlower}) detect the exact evaluation of the Gini index. 

We should notice that both upper and lower bounds can be improved easily
by considering a `polygon' surrounding or including the area of a half Gini index $g/2$.
Using the procedure, one can improve the bounds recursively and systematically.
Eventually, the both bounds are expected to be very closed to the true $g$. 
\section{Results for mixture of distributions}
\label{sec:results}
In the previous section, we introduced Gini and $k$ indices and 
discuss the generic properties. From the definition of these measures, 
we always evaluated the values empirically from a finite number of data set. 
However, as we showed in Sec.~\ref{sec:general}, it is easy to calculate the value analytically when the distribution 
of population is described by parametric distribution such as 
a uniform, power law and lognormal distributions. 
In fact, in the previous section, we derived the measures for these distribution functions. 
Turning now to the situation of reality, 
the distribution of population like wealth, number of citation, etc. is 
well fitted to a mixture of uniform, power-law and lognormal distributions. 
\subsection{Empirical data}
\label{subsec:empirical}
In this paper, we focus on three types of socio-economic data: 
(i) voting 
(ii) citations of different science journals, and
(iii) population of cities and municipalities.

We use the voting data for open-list proportional elections from several countries of Europe 
(data taken from Ref.~\cite{chatterjee2013universality}).
The number of votes $v_i$ of a candidate is divided by the average number of votes $v_0$  
of all candidates in his/her party list. 
We focus on the probability distribution $P(v/v_0)$ of the  quantity $v/v_0$, known as the `performance'
of a candidate. We use data for Italy, Netherlands and Sweden.

For citations to journals, we collected data from ISI Web of Science~\cite{ISI}, 
citations gathered until a certain date by all articles/papers published in a particular year, for a (i) few scientific journals
(PRL = Physical Review Letters, CPL = Chemical Physics Letters, PRA = Physical Review A, PNAS = Proceedings of the National Academy of Sciences USA,
Lancet, BMJ = British Medical Journal, NEJM = New England Journal of Medicine),
and (ii) universities/institutions (University of Oxford,  University of Cambridge,  University  of Tokyo, University of Melbourne).
We computed the  probability distribution of citations $p(c)$, and found the corresponding scaling collapses for
similar categories, by rescaling with the average number of citations $\langle c \rangle$.

We also collected data for city sizes for Brasil~\cite{Brasil}, municipalities of Spain~\cite{Spain} and Japan~\cite{Japan}.
We computed the  probability distribution of city/ municipality population $p(s)$, and rescaled them 
with the average population $\langle s \rangle$.

\begin{figure}[ht]
 \includegraphics[width=17.9cm]{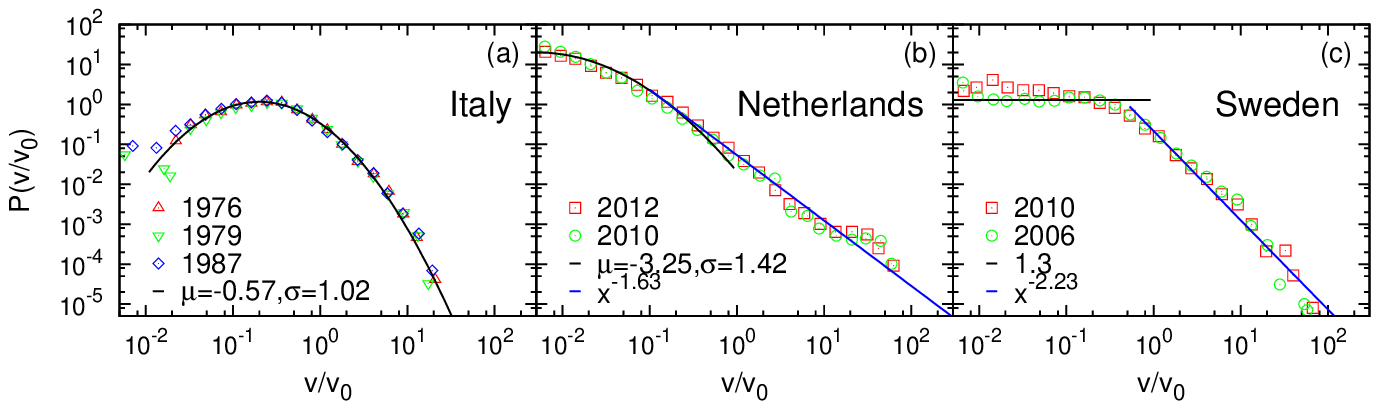} 
\includegraphics[width=17.9cm]{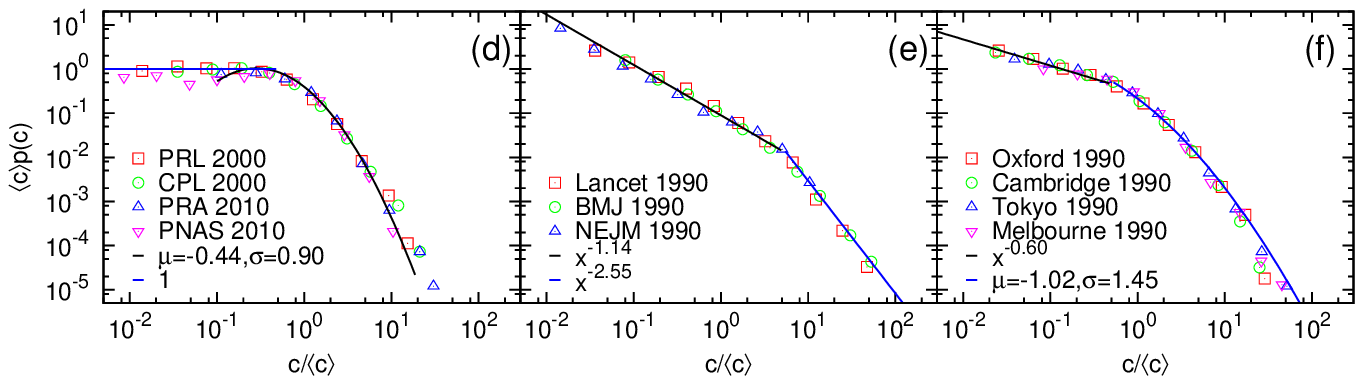}
\includegraphics[width=17.9cm]{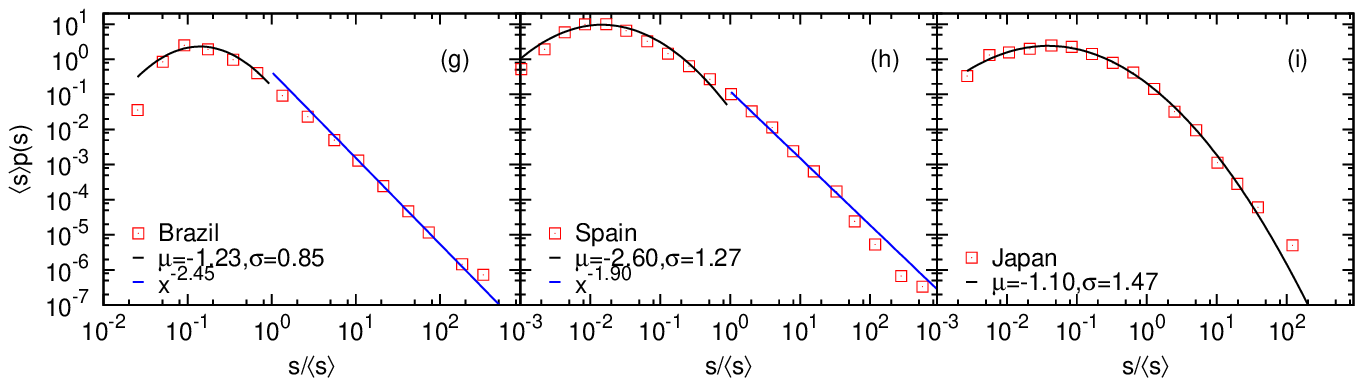}
\caption{Figure showing broad distributions of different quantities and their fitting.
(a) Distribution of performance ($v/v_0$) for candidates in open list proportional elections in Italy for several years, 
fitted to lognormal function. Data taken from Ref.~\cite{chatterjee2013universality}.
(b) for proportional elections with semi-open lists in Netherlands, fitted to lognormal followed by  power law distribution.
(c) for proportional elections with semi-open lists in Sweden, fitted to uniform followed by  power law distribution.
(d) Distribution of citations for journals, fitted to uniform distribution followed by lognormal~\cite{ISI}.
(e) double power law.
(f) power law, followed by lognormal. 
Population of
(g) cities in Brazil~\cite{Brasil}, the same as case (b),
(h) municipalities of Spain~\cite{Spain}, the same as case (b),
(i) municipalities of Japan~\cite{Japan}, the same as (a). 
}
\label{fig:various1}
\end{figure}

In Fig.~\ref{fig:various1}, we show broad distributions of the above quantities and their fitting.
From this Figure, we are confirmed that 
the distribution which is well-fitted to these empirical evidences 
falls into six categories, 
namely, 
(a) a single lognormal, 
(b) a single lognormal with a power law tail, 
(c) uniform with a power law tail, 
(d) uniform with a lognormal tail, 
(e) a mixture of power laws, 
(f)  a single power law with a lognormal tail. 

Therefore, it is worth while for us to 
prepare the formula to calculate the measures analytically for those six cases. 

\subsection{Analytic formulas for six categories and relationship between $g$ and $k$}
We plot the typical behavior 
of Lorentz curve, $g$ and $k$ indexes by the explicit formulas for 
the cases of (b)-(f) in Fig.~\ref{fig:fg_five_cases}. 
The formulas and the details are given in Appendix \ref{app:ap_der_general}. 

In practice, 
it is convenient for us to clarify the relationship between 
$g$ and $k$ indices. 
As we already showed, 
these two indices are both dependent on 
the crossover point $m_{\times}$ as 
$g(m_{\times})$ and $k(m_{\times})$. 
It is difficult for us to obtain the relation 
analytically. However, one can obtain it numerically 
when we consider $m_{\times}$ as `time' and also 
consider the `trajectory' in the $g$-$k$ space. 
\begin{figure}[t]
\begin{center}
\includegraphics[width=8.9cm]{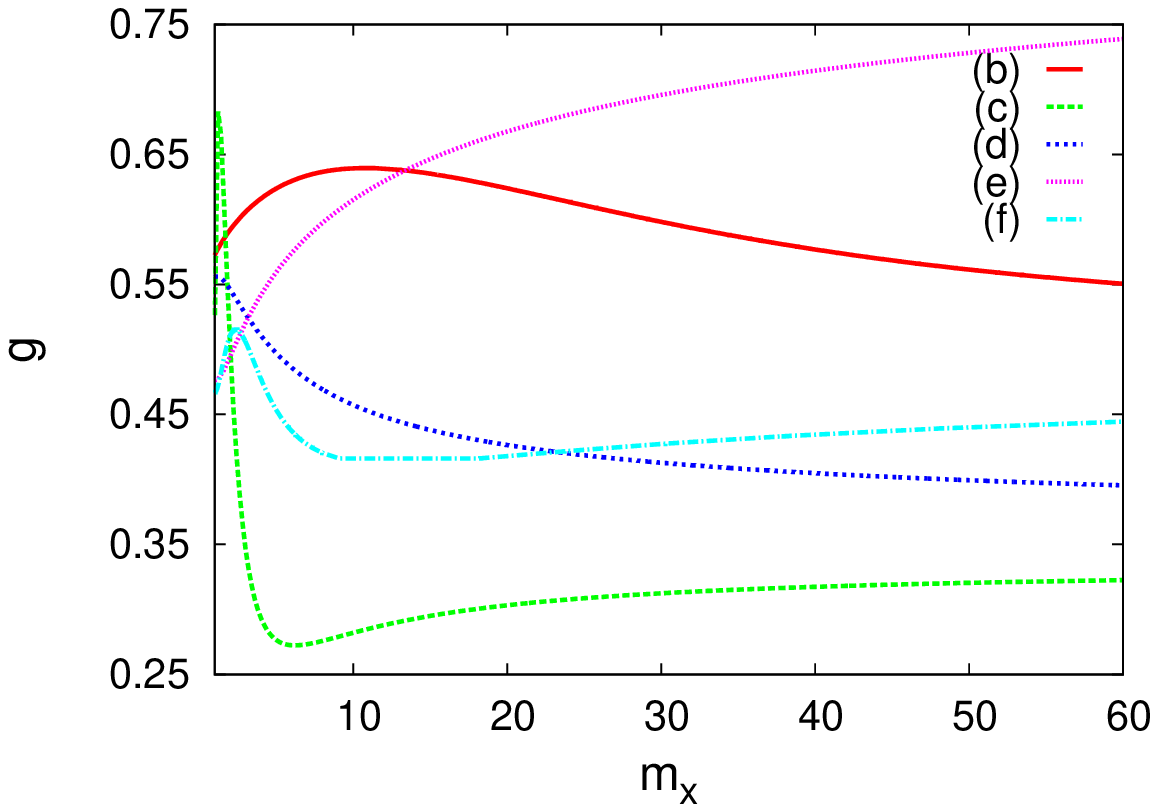}
\includegraphics[width=8.9cm]{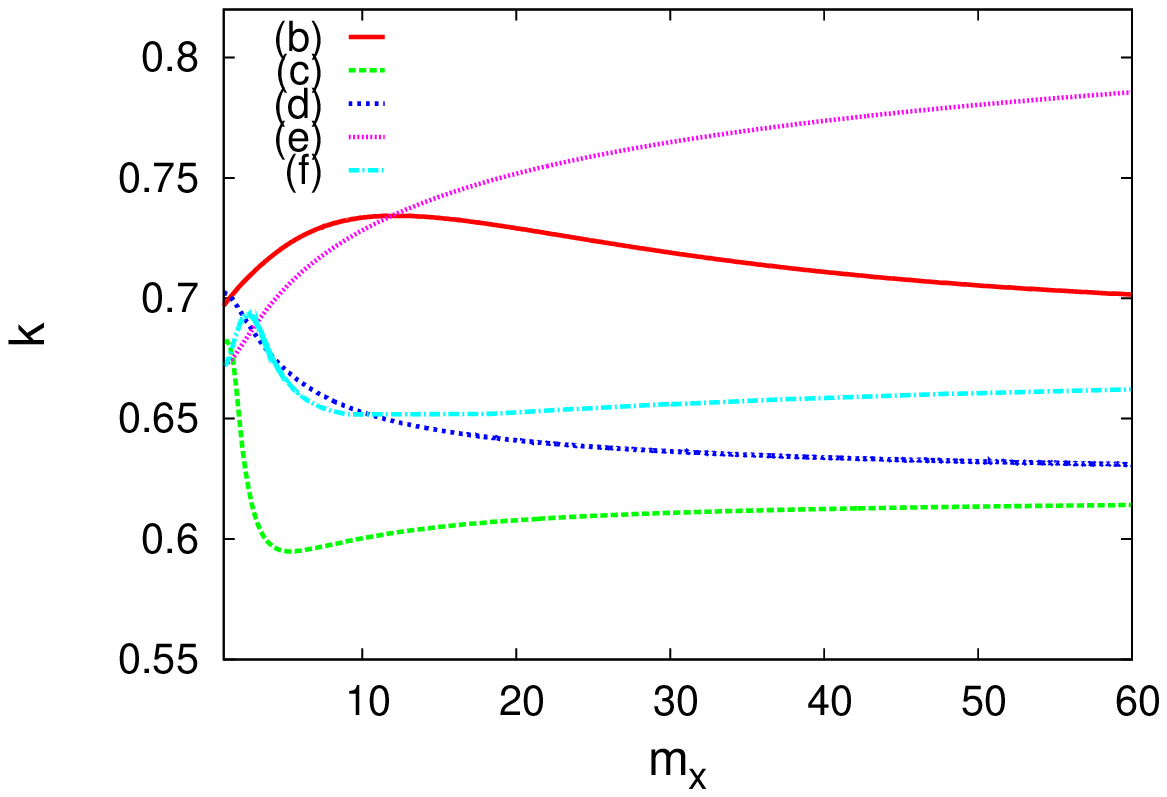}
\includegraphics[width=8.9cm]{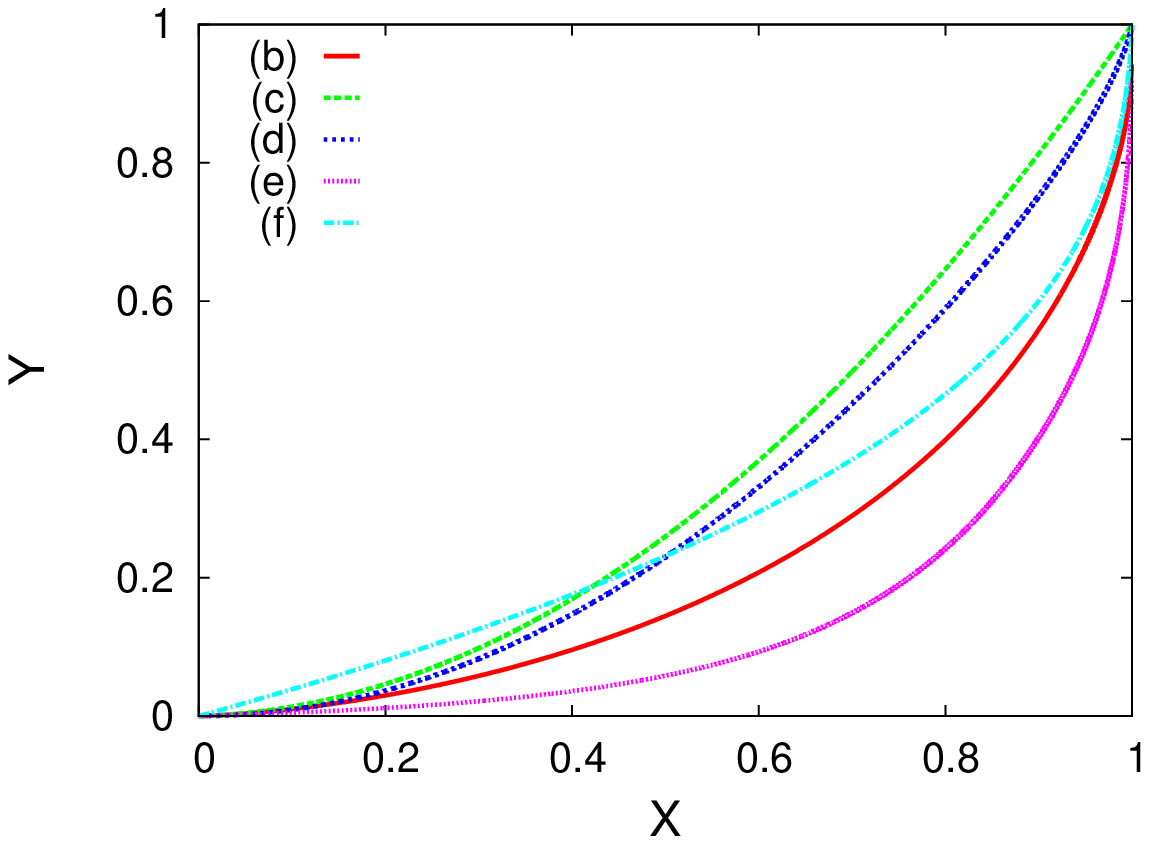}
\includegraphics[width=8.9cm]{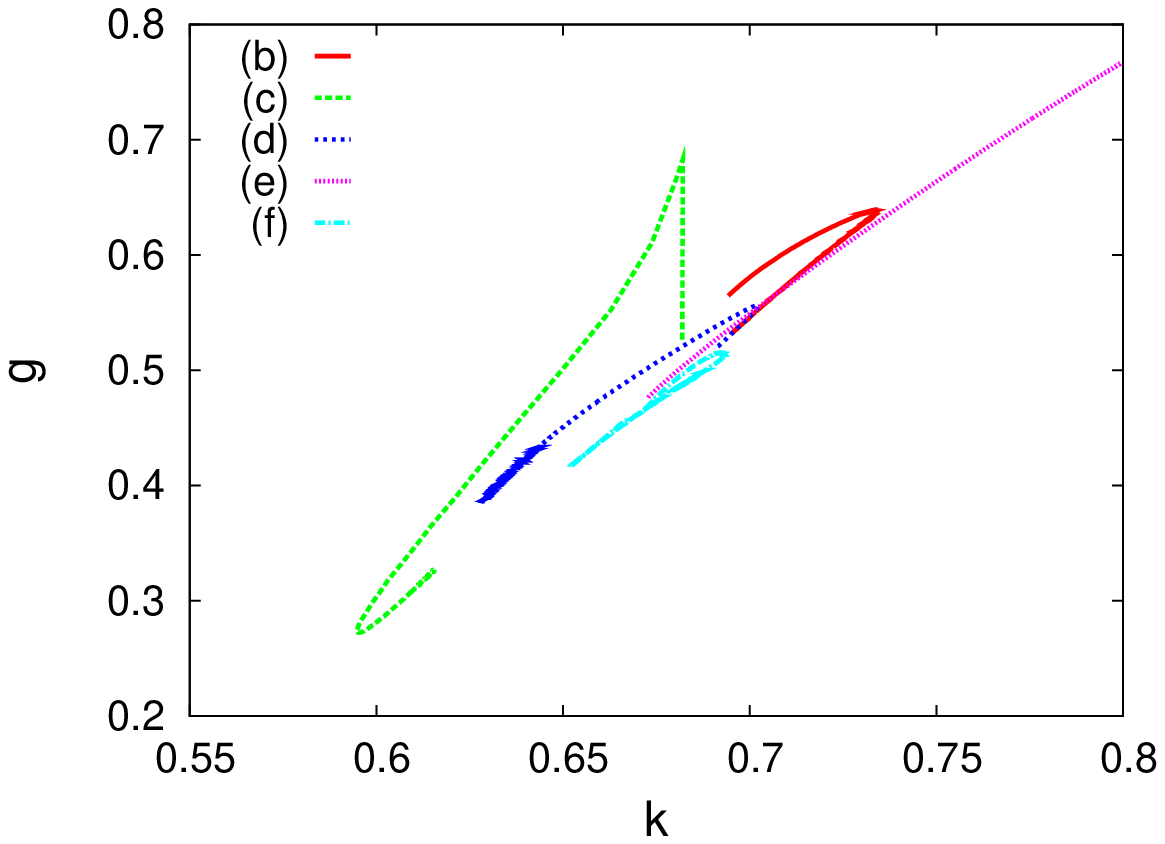}
\end{center}
\caption{
Clockwise from the upper left, 
we show the $m_{\times}$-dependence of the $g$ and $k$ indices, 
the relationship between 
$g$ and $k$ indices 
for five cases (b)-(f) (refer to Fig.~\ref{fig:various1}), and the Lorentz curve. 
We set 
$\sigma=\mu=1$ for lognormal distributions appearing in (b), (d), (f) and 
$\alpha=2.5$ for power law distribution in (b), (c). 
In the case (c), we set $a=1$ as a constant for uniform distribution. 
In the case (e), the two exponents for power law distributions are set to 
$\alpha=1.14$ and $\beta=2.55$. 
}
\label{fig:fg_five_cases}
\end{figure}
In Fig.~\ref{fig:fg_five_cases} (lower right), 
we show the trajectories 
for five cases. 
From this panel, we confirmed that 
$g$ as a function of $k$ 
is single-valued function only for the case (e) 
for the parameters taken in these plots. 
We should also notice that 
the Gini index $g$  changes  almost linearly as a function of $k$. 
Hence, one can infer the value of $g$ 
by using the relationship when the parameter sets including 
$m_{\times}$ are known beforehand. 
On the other hand, for the other cases, 
$g$ are piecewise multi-valued functions. 
Therefore, we should keep in mind that 
for a given value of $k$, there are several candidates 
for the corresponding $g$ value depending on 
the crossover point $m_{\times}$. 
\subsection{Gap between empirical and analytical values of measures}
To compare the empirical and 
analytical estimates for $g$ and $k$ indices whose 
distribution functions are shown in 
Fig.~\ref{fig:various1}, 
we need to calculate those measures from empirical data sets. 
It is important for us to bear in mind that 
the measures could be calculated independently from 
the analytic formulas from data sets.
 
Let us consider the data set of 
wealth for $N$ persons:  
$m_{1} \leq m_{2} \leq \cdots \leq m_{N}$. 
Then, the Lorenz curve is 
given by 
\begin{equation}
X(r ) = 
\frac{N}{r},\,\,\,
Y(r ) = 
\frac{\sum_{i=1}^{r}m_{i}}
{\sum_{i=1}^{N}m_{i}}=\frac{1}{\mu N}\sum_{i=1}^{r}m_{i},
\label{eq:empirical_L}
\end{equation}
where we defined the empirical mean $\mu=(1/N)\sum_{i=1}^{N}m_{i}$. 
Using the general 
definition of the Gini index Eq.~(\ref{eq:def_G}), 
one can obtain the  $g$ 
using the following non-parametric way as 
\begin{equation}
\hat{g}=
\frac{2}{\mu N^{2}}
\sum_{r=1}^{N}
r m_{r}-\frac{(N+1)}{N}. 
\label{eq:empirical_G}
\end{equation}
The detail of derivation is explained in Appendix \ref{app:ap_der_Gini}. 

On the other hand, empirical $k$ index, say $\hat{k}$ is given by 
\begin{equation}
\hat{k} = \frac{r_{*}}{N}, 
\end{equation}
where $r_{*}$ is 
the solution of 
the following equation
\begin{equation}
\frac{r_{*}}{N}=1-
\frac{\sum_{i=1}^{r_{*}}
m_{i}}
{\mu N}.
\end{equation}
Of course, it is hard to find the solution with precision 
for finite $N$, we might use in practice 
\begin{equation}
r_{*}=
\arg\min_{r}
\left|
\frac{r}{N}-
1+
\frac{\sum_{i=1}^{r}m_{i}}{\mu N}
\right|. 
\end{equation}
\begin{table}[t]
\caption{Estimates of Gini ($g$) and $k$ index for distributions of different quantities: 
voting data (Taken from Ref.~\cite{chatterjee2013universality}), citation data for journals and institutions~\cite{ISI}
and sizes of cities and municipalities~\cite{Brasil,Spain,Japan}.
The functional fits and their ranges are mentioned. 
where $\hat{g},\hat{k}$ are 
empirical estimates while $g$ and $k$ are computed from the analytical functions.
$N$ is the number of data and 
$\langle \cdot \rangle$ denotes 
the empirical mean 
$\mu = (1/N)\sum_{i=1}^{N}m_{i}$. 
}
\begin{tabular}{ |c|c|c|c|c|c|c|c|c|c|c|}
\hline

Country  & Year & $\hat{g} $ &  $g$ & $\hat{k}$ & $k$ & $m_{\times} $ & $m<m_{\times}$ & $m>m_{\times} $ &  $N$  & $v_{0}$\\ 

\hline 

\multirow{4}{*}{Italy}   & 1976 & 0.5593 &  &  0.7077 &  &  & log-normal & & 5839 & 1.0 \\   
& 1979 & 0.5463 & 0.5292 & 0.7014 & 0.6948 & - & $\mu=-0.57; \sigma=1.02$ & - & 7153 & 1.0 \\   
& 1987 & 0.5720 & & 0.7144 &  & &  & & 8620 & 1.0\\ 

\hline 

\multirow{2}{*}{Netherlands} & 2010 &  0.9406 &   & 0.9214 &  &  & log-normal & power law & 7229 & 1.0 \\   
& 2012 & 0.9250 & 0.8038 & 0.9071 & 0.8935 & 0.20 & $\mu=-3.25; \sigma=1.42$ & $\alpha=1.64$ & 8889 & 1.0 \\   

\hline 

\multirow{2}{*}{Sweden} & 2006  & 0.6903 & & 0.7650 &  &  & uniform & power law & 5150 & 1.0\\   
&  2010 & 0.7374 & 0.6825 & 0.7842 & 0.7315 & 0.50 & $1/a=1.3$ & $\alpha=2.24$ & 9053 & 1.0 \\   
\hline 
\end{tabular}
\mbox{}\\
\mbox{}\\ 
\mbox{}\\
\begin{tabular}{ |c|c|c|c|c|c|c|c|c|c|c|}
\hline

Journals/  & Year & $\hat{g}$ & $g$ &  $\hat{k}$ & $k$ & $m_{\times} $ & $m<m_{\times}$ & $m>m_{\times} $ & $N$ & $\langle c \rangle $\\ 
Institutions   &  &  &  & & & & & & &\\ 

\hline 
\multirow{1}{*}{PRL} & 2000 & 0.5859 &  & 0.7154  & & & & & 3124 & 72.21 \\   

\multirow{1}{*}{CPL} & 2000 & 0.5788  & & 0.7123 &  & & uniform & log-normal & 1512  & 28.44 \\   
\multirow{1}{*}{PRA} & 2010 & 0.5271  & 0.5214 & 0.6895 & 0.6880 & 0.30 & $1/a=0.95$ & $\mu=-0.44; \sigma=0.90$ & 1410  & 27.62 \\   

\multirow{1}{*}{PNAS} & 2010 & 0.4616  & & 0.6641  & & & &  & 2698  & 117.01 \\   
\hline 
\multirow{1}{*}{Lancet} &1990 & 0.8448  & & 0.8410   & & & power law  &  power law & 3232  & 27.71\\   

\multirow{1}{*}{BMJ} & 1990 & 0.8840 & 0.8808 & 0.8662 & 0.8660 & 5.0 & $\alpha=1.14$ &  $\beta=2.55$ & 2847  & 12.48 \\   
\multirow{1}{*}{NEJM} & 1990 & 0.8536  & & 0.8498   & & & & & 1684  & 69.28\\   

\hline 
\multirow{1}{*}{Oxford} & 1990 & 0.7276  & & 0.7769  & & & &  & 2147  & 39.10 \\   
\multirow{1}{*}{Cambridge} &  1990  & 0.7366  & & 0.7791  & & & power law &  log-normal & 2616  & 42.74\\   

\multirow{1}{*}{Tokyo} & 1990 & 0.6834  & 0.4755 & 0.7564 & 0.6664 & 0.50 & $\alpha=0.60$& $\mu=-1.02; \sigma=1.46$ & 4196  & 25.77\\   

\multirow{1}{*}{Melbourne} & 1990 & 0.6772  & & 0.7515  & & & & & 1131  & 26.83 \\   
\hline 
\end{tabular}
\mbox{}\\
\mbox{}\\
\mbox{}\\
\begin{tabular}{ |c|c|c|c|c|c|c|c|c|c|c| }
\hline

Country $~~$  & Year & $\hat{g}$ &  $g$ & $\hat{k}$ & $k$ & $m_{\times}$ & $m<m_{\times}$ & $m>m_{\times}$  & $N$ & $\langle s \rangle $ \\ 
\hline 
  &   &  &  & & & & & & & \\ 
Brazil   & 2012  & 0.7270 & 0.6253  & 0.7795 & 0.7275 & 1.0 & log-normal & power law & 5570 & 34825.23 \\ 
  &  &  &  &  & & & $\mu=-1.23; \sigma=0.86$ & $\alpha=2.45$ & &\\

\hline 
 &   &  &  & & & & & & & \\ 
Spain   & 2011  &  0.8661 & 0.8451 & 0.8560  & 0.8469 & 1.0 & log-normal & power law & 8116 & 5814.50\\ 
 &   &  &  & & & & $\mu=-2.60; \sigma=1.27$ & $\alpha=1.90$ & &\\ 
\hline 
 &   &  &  & & &  & & & & \\ 
Japan   & 2010 &  0.7192 &  0.7014 & 0.7738  & 0.7689 & - & log-normal & -& 1720 & 74451.95\\ 
&   &  &  & & & & $\mu=-1.10; \sigma=1.47$ & & &\\ 
\hline 

\end{tabular}
\label{tab:fits}
\end{table}

From the Table \ref{tab:fits}, we found that there is a finite gap 
between empirical and analytical values of measures. 
When we assume that 
the exponent 
such as $\alpha, \beta, \mu, \sigma$ are precisely determined 
by means of maximum likelihood estimate, 
one can estimate the crossover point $m_{\times}$ by minimizing 
the gaps $\Delta g(m_{\times}), \Delta k (m_{\times})$. 
Namely, the cost function for the estimation could be constructed by 
\begin{equation}
\Delta_{g} (m_{\times}) = 
(\hat{g}-g(m_{\times}))^{2},\quad 
\Delta_{k} (m_{\times}) = 
(\hat{k}-k(m_{\times}))^{2},
\end{equation}
where $\hat{g}$, $\hat{k}$ are 
empirical estimates, whereas 
$g(m_{\times})$, $k(m_{\times})$ are 
the analytical expressions as a function of 
the crossover point $m_{\times}$. 
\begin{figure}[t]
\begin{center}
\includegraphics[width=8.9cm]{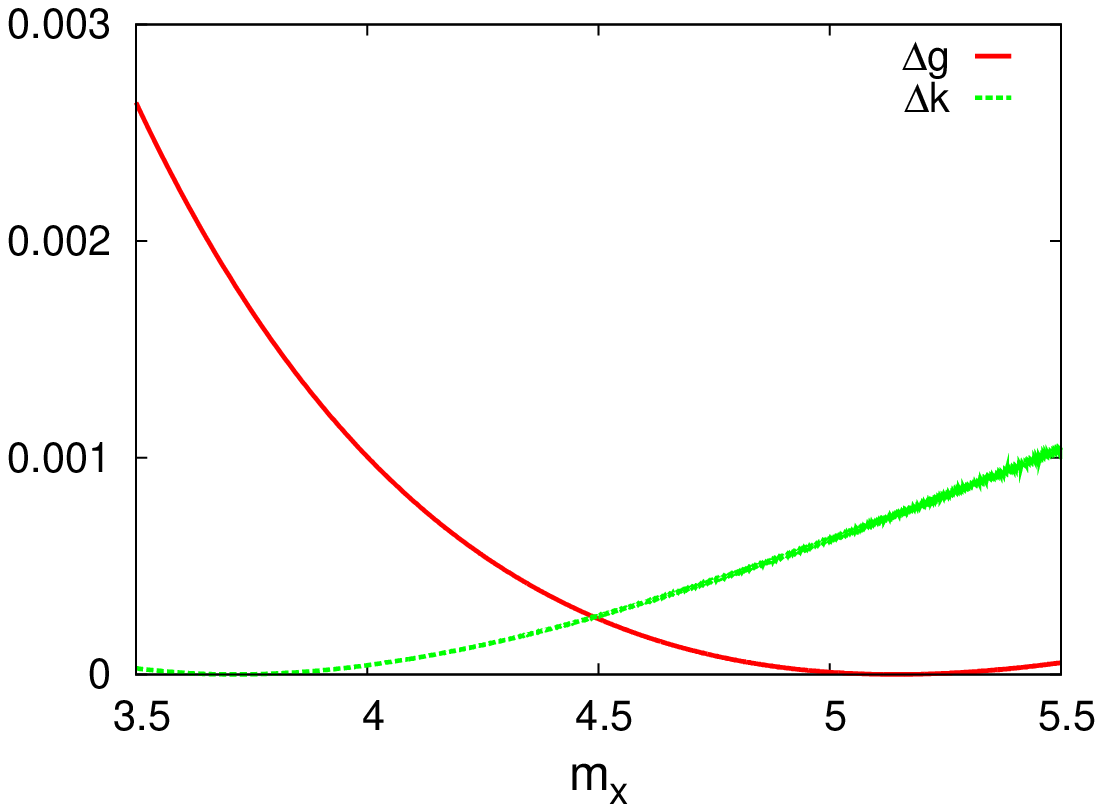}
\includegraphics[width=8.9cm]{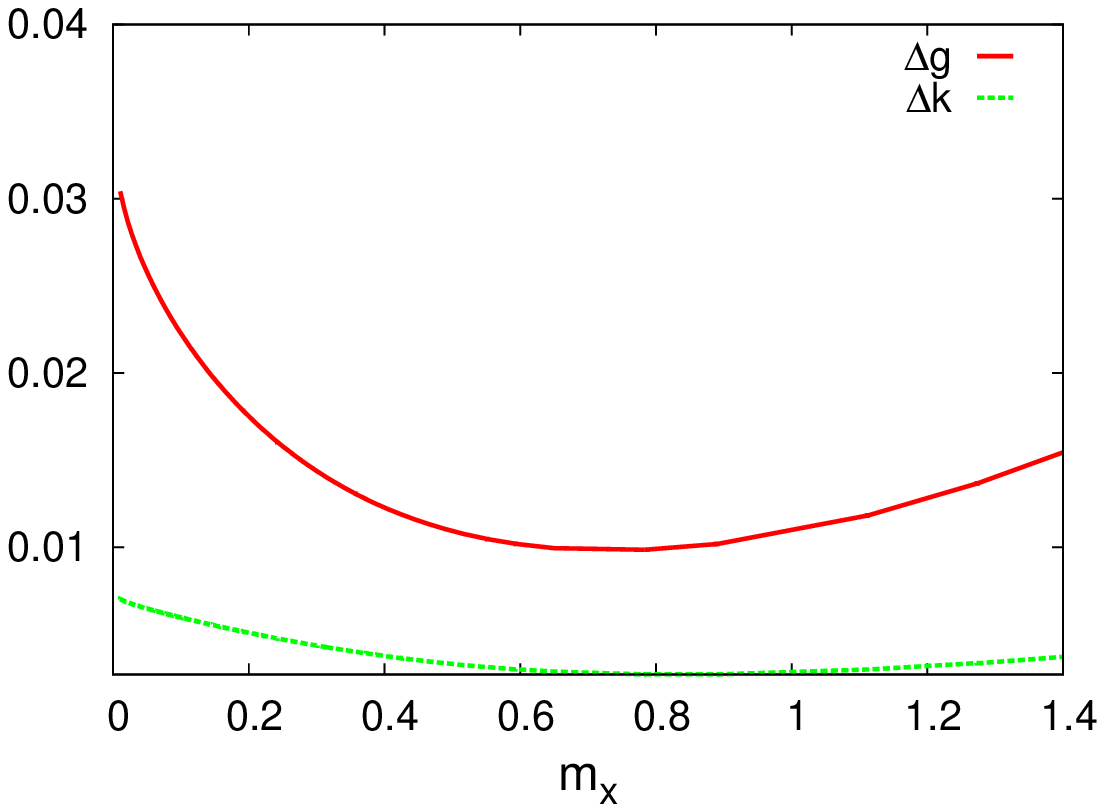}
\end{center}
\caption{
The gaps $\Delta g, \Delta k$ as a function of the crossover point $m_{\times}$ for 
the case of (e) (left) and (g) (right). 
For (e), we chose 
$\hat{g}, \hat{k}$ of 
`Lancet' because $N$ is largest among three of 
(Lancet, BMJ, NEJM). 
}
\label{fig:fg_gap}
\end{figure}
We show the result in Fig.~\ref{fig:fg_gap} for the case of (e) (left) and (g) (right).  
Then minimum point is estimate by means of minimization of the gap. 
For the case of (e), we find $m_{\times}=5.16$ for minimization of $\Delta g$ and 
$m_{\times}=3.74$ for minimization of $\Delta k$, and the resulting 
values of $g$ and $k$ are 
$g=0.8844$ and $k=0.8416$, respectively.  
On the other hand, 
for the case of (g), we have 
$m_{\times}=0.765$ for minimization of $\Delta g$ and 
$m_{\times}=0.767$ for minimization of $\Delta k$, 
which lead to 
the corrected estimates $g=0.6297$ and $k=0.7281$. 
\section{Summary and discussions}
\label{sec:summary}

The probability distributions of several socio-economic quantities showing inequality have broad distributions.
In Sec.~\ref{sec:Generic}, we presented the general form of 
the inequality measures, 
by computing the Lorenz curve, and using that to compute the 
Gini and $k$ indices for a class of distributions.
From the empirical data we analyzed, 
we showed that the distributions can be put into 
several categories and each of these can be specified by an appropriate 
parametric distribution. In fact, we found six categories of  distributions,
most of which are a mixture of two distinct distributions with a crossover point. 
In Sec.~\ref{sec:general}, 
we computed the general formulas of the inequality measures  
for the combinations of functions as observed from empirical distributions,  
and compared them with those from analytical calculations.
In Sec.~\ref{sec:inequality}, we considered 
the $k$ index as an `extra dimension', both the lower and upper bounds of the Gini index are  obtained 
as a function of the $k$ index. This type of inequality relation between inequality indices
might help us to check the validity of 
empirical and analytical evaluations of these indices. 
In Sec.~\ref{sec:results}, we reported our results.  
We provided numerical evaluations of our measures, and 
compared their empirical and analytical values.
By minimizing the gap between two results obtained in 
different ways, we provide the estimates of the best possible crossover point 
for a given data set.

Socio-economic inequality is a topic of major concern~\cite{Sciencespecialissue}, 
drawing attention of researchers across various disciplines. 
Researchers have always concentrated on (i) characterizing empirical data
and thereby computing inequality measures like Gini index, and (ii) modeling
the origins of broad distributions.
Our paper focuses on extensively computing the Gini index $g$ and the newly introduced $k$ index
from single analytical distributions or combinations of them, which fit well to
empirical data, and also compare the results with those calculated directly
from the empirical data sets themselves.
Our proposed quantitative methodology to estimate the crossover point between two functional fits
to empirical data could also prove to be useful, beyond the realm of inequality research.
While the much studied Gini index gives an overall measure
of the inequality, the $k$ index tells us that the cumulative wealth of $(1-k)$ fraction
of individuals are held by $k$ fraction of individuals.

\subsection*{Acknowledgement}
J.I. was financially supported by Grant-in-Aid for Scientific Research (C) of
Japan Society for the Promotion of Science (JSPS) No. 2533027803 and 
Grant-in-Aid for Scientific Research (B) of 26282089, 
Grant-in-Aid for Scientific Research on Innovative Area No. 2512001313. 
He also thanks Saha Institute of Nuclear Physics for their hospitality 
during his stay in Kolkata.
B.K.C. and A.C. acknowledges support from B.K.C.'s J.~C.~Bose Fellowship and  Research Grant.

\appendix 
\section{Derivation of the general forms of inequality measures}
\label{app:ap_der_general}
Here we drive the general form Eq.~(\ref{eq:general_L})(\ref{eq:general_G}) and Eq.~(\ref{eq:general_k}). 
for the mixture of two normalized distribution Eq.~(\ref{eq:Pm}). 
From the definition of Lorenz curve, we have 
\begin{eqnarray}
X(r ) & = & 
\frac{Q_{1}(r )}
{Q_{1}(m_{\times})+Q_{2}(m_{\times})}
\theta (r,m_{\times})+
\left\{
1-
\frac{Q_{2}(r )}{Q_{1}(m_{\times})+Q_{2}(m_{\times})}
\right\} \Theta (r-m_{\times}) \\
Y (r ) & = & 
\frac{R_{2}(r )}
{R_{1}(m_{\times})+R_{2}(m_{\times})}\theta (r,m_{\times})
+
\left\{
1-
\frac{R_{2}(r )}{R_{1}(m_{\times})+R_{2}(m_{\times})}
\right\} \Theta (r-m_{\times})
\end{eqnarray}
For $r$ of active $\theta (r,m_{\times})=1$, we have $(Q_{1}(m_{\times})+Q_{2}(m_{\times}))X=Q_{1}( r)$, 
namely, $r=Q_{1}^{-1}[Q_{1}((m_{\times})+Q_{2}(m_{\times}))X]$, and this reads 
\begin{equation}
Y = \frac{R_{1}(Q_{1}^{-1}(Q_{1}(m_{\times})+Q_{2}(m_{\times}))X)}
{R_{1}(m_{\times})+R_{2}(m_{\times})}, \quad \quad 
0 \leq X \leq 
\frac{Q_{1}(m_{\times})}
{Q_{1}(m_{\times})+Q_{2}(m_{\times})}.
\label{eq:der_L1}
\end{equation}
On the other hand, 
for $r$ of active $\Theta (r-m_{\times})=1$, we have 
$(1-X)(Q_{1}((m_{\times})+Q_{2}(m_{\times})))=Q_{2} (r )$, namely,
$r=Q_{2}^{-1}
[(Q_{1}((m_{\times})+Q_{2}(m_{\times})))(1-X)]$ and $Y$ is given by 
\begin{equation}
Y = 
 1-
 \frac{R_{2}^{-1}[Q_{2}^{-1}
[(Q_{1}((m_{\times})+Q_{2}(m_{\times})))(1-X)]]}
{R_{1}(m_{\times})+R_{2}(m_{\times})},\quad \quad 
\frac{Q_{1}(m_{\times})}
{Q_{1}(m_{\times})+Q_{2}(m_{\times})} <X \leq 1.
\label{eq:der_L2}
\end{equation}
Therefore, Eq.~(\ref{eq:der_L1}) and Eq.~(\ref{eq:der_L2}) are the general form of the Lorenz curve Eq.~(\ref{eq:general_L}).  

The Gini index is calculated by 
\begin{eqnarray}
G & = & 
2\int_{m_{0}}^{\infty}(X(r )-Y(r ))\frac{dX( r)}{dr}dr \nonumber \\
\mbox{} & = & 
\frac{2\int_{m_{0}}^{m_{\times}}
Q_{1}(r ) \frac{dQ_{1} (r )}{dr}dr}{
\{Q_{1}(m_{\times})+Q_{2}(m_{\times})\}^{2}}
 -
\frac{2\int_{m_{0}}^{m_{\times}}
R_{1}(r ) \frac{dQ_{1} ( r)}{dr}dr}
{
\{Q_{1}(m_{\times})+Q_{2}(m_{\times})\}
\{R_{1}(m_{\times})+R_{2}(m_{\times})\}
} \nonumber \\
\mbox{} &  & 
+ \frac{2\int_{m_{\times}}^{\infty}
Q_{2}(r ) \frac{dQ_{2} (r )}{dr}dr}{
\{Q_{1}(m_{\times})+Q_{2}(m_{\times})\}^{2}}
 -
\frac{2\int_{m_{\times}}^{\infty}
R_{2}(r ) \frac{dQ_{2} ( r)}{dr}dr}
{
\{Q_{1}(m_{\times})+Q_{2}(m_{\times})\}
\{R_{1}(m_{\times})+R_{2}(m_{\times})\}
}. 
\end{eqnarray}
When we notice 
\begin{eqnarray}
\int_{m_{0}}^{m_{\times}}
Q_{1}(r ) \frac{dQ_{1} (r )}{dr}dr & = & 
\int_{m_{\times}}^{\infty}\frac{1}{2}\frac{d}{dr}\{Q_{1}(r )^{2}\}dr=\frac{1}{2}\{Q_{1}(m_{\times})^{2}-Q_{1}(m_{0})^{2}\} \\
{\rm and} \int_{m_{\times}}^{\infty}
Q_{2}(r ) \frac{dQ_{2} (r )}{dr}dr & = & 
\int_{m_{\times}}^{\infty}
\frac{1}{2}\frac{d}{dr}
\{Q_{2}(r )^{2}\}dr = \frac{1}{2}\{Q_{2}(\infty)^{2}-Q_{2}(m_{\times})^{2}\},
\end{eqnarray}
and using the definition of 
$S_{1}(m_{0},m_{\times})$ and $T_{2}(m_{\times})$ (see Eq.~(\ref{eq:def_S1T2})), we obtain the general form Eq.~(\ref{eq:general_G}). 

The general form of the $k$ index Eq.~(\ref{eq:general_k}) is simply obtain 
by setting $X=k$ and $Y=1-k$, which means $Y=1-X$,  in Eq.~(\ref{eq:der_L1}) and Eq.~(\ref{eq:der_L2}). 
\section{Derivation of empirical Gini index}
\label{app:ap_der_Gini}
Here we show the derivation of empirical form of the Gini index Eq.~(\ref{eq:empirical_G}). 
From the discrete expressions Eq.~(\ref{eq:empirical_L}) with the relation $dX=(r+1)/N-rN=1/N$, the Gini index 
$\hat{g}$ is written by 
\begin{eqnarray}
\hat{g}  & = & 
2\sum_{r=1}^{N}(X_{r}-Y_{r})\frac{1}{N} \nonumber \\
\mbox{} & = & 
\frac{2}{N^{2}}\sum_{r=1}^{N}r-\frac{2}{\mu N^{2}}\sum_{r=1}^{N}\sum_{i=1}^{r}m_{i} \nonumber \\
\mbox{} & = & 
\frac{(N+1)}{N}
-
\frac{2}{\mu N^{2}}
\sum_{r=1}^{N}(N-r+1)m_{r} \nonumber \\
\mbox{} & = & 
\frac{(N+1)}{N}
-\frac{2}{\mu N^{2}}
\left\{
\mu N(N+1)-\sum_{r=1}^{N}r m_{r}
\right\} \nonumber \\
\mbox{} & = & 
\frac{2}{\mu N^{2}}
\sum_{r=1}^{N}r m_{r}-
\frac{(N+1)}{N}. 
\end{eqnarray}
This is nothing but Eq.~(\ref{eq:empirical_G}). 

\section{Explicit forms of measures for mixture of distributions}
\label{app:ap_der_measures}
\subsection{(b) Lognormal with a power-law} 
We consider the case of 
$F_{1}(m)={\rm e}^{-\frac{(\log m-\mu)^{2}}{2\sigma^{2}}}/\sqrt{2\pi}\sigma m$ and 
$F_{2}(m)=(\alpha-1)m^{-\alpha}$ with $m_{0}=0$. 
We have 
\begin{eqnarray}
Q_{1}(r ) & = &  H\left(
\frac{\mu -\log r}{\sigma}
\right),\,\,\,
Q_{2}(r )=r^{1-\alpha};  \\
R_{1}(r ) & = & 
{\rm e}^{\mu + \frac{\sigma^{2}}{2}}
H
\left(
\frac{\mu + \sigma^{2}-\log r}{\sigma}
\right),\,\,\,
R_{2}(r ) = 
\frac{(\alpha-1)r^{2-\alpha}}{(\alpha-2)}. 
\end{eqnarray}
Then, using the following staffs 
$Q_{1}(m_{\times})=H(\frac{\mu -\log m_{\times}}{\sigma}), 
Q_{1}(m_{0})=Q_{2}(\infty)=0, 
Q_{2}(m_{\times})=m_{\times}^{1-\alpha}$, 
and $R_{1}(m_{\times})={\rm e}^{\mu + \frac{\sigma^{2}}{2}}H(\frac{\mu + \sigma^{2}-\log m_{\times}}{\sigma}), 
R_{2}(m_{\times})=(\alpha-1)m_{\times}/(\alpha-2)$,  
and accompanying 
\begin{eqnarray}
S_{1}(m_{0},m_{\times}) & = & 
{\rm e}^{\mu + \frac{\sigma^{2}}{2}}
\int_{\frac{\mu + \sigma^{2}-\log m_{\times}}{\sigma}}^{\infty}Dx H(x+\sigma),  \\
T_{2}(m_{\times}) & = & 
-\frac{(\alpha-1)^{2}m_{\times}^{3-2\alpha}}{(\alpha-2)(2\alpha-3)}, 
\end{eqnarray}
the Lorenz curve is given as 
\begin{equation}
Y = 
\left\{
\begin{array}{lc}
\frac{{\rm e}^{\mu +\frac{\sigma^{2}}{2}}
H
(
\sigma -
H^{-1}
[
\{
H
(
\frac{\mu-
\log m_{\times}^{1-\alpha}}
{\sigma}
)+m_{\times}^{1-\alpha}
\}X
]
)
}
{{\rm e}^{\mu + \frac{\sigma^{2}}{2}}
H
(
\frac{\mu + \sigma^{2}-\log m_{\times}}
{\sigma}
)
+
\left(
\frac{\alpha-1}{\alpha-2}
\right)
m_{\times}^{1-\alpha}
}, & 
\quad \quad 
0 \leq X \leq 
\frac{H(\frac{\mu -\log m_{\times}}{\sigma})}
{H(\frac{\mu-\log m_{\times}}{\sigma})+
m_{\times}^{1-\alpha}},
\\
1-
\frac{
\left(
\frac{\alpha-1}{\alpha-2}
\right)
[
H(\frac{\mu \log m_{\times}}{\sigma})
+m_{\times}^{1-\alpha}]^{\frac{2-\alpha}{1-\alpha}}
}
{ 
{\rm e}^{\mu + \frac{\sigma^{2}}{2}}
H(\frac{\mu + \sigma^{2}-\log m_{\times}}{\sigma})
+
\left(
\frac{\alpha-1}{\alpha-2}
\right)
m_{\times}^{2-\alpha}
}
(1-X)^{\frac{2-\alpha}{1-\alpha}},
 & 
\quad \quad 
\frac{H(\frac{\mu -\log m_{\times}}{\sigma})}
{H(\frac{\mu-\log m_{\times}}{\sigma})+
m_{\times}^{1-\alpha}}
< X \leq 1,
\end{array}
\right. 
\end{equation}
and the $g$ index is 
\begin{equation}
g =  
\frac{
H(\frac{\mu -\log m_{\times}}{\sigma})^{2}
-m_{\times}^{2-2\alpha}}
{
(
H(\frac{\mu -\log m_{\times}}{\sigma}) + 
m_{\times}^{1-\alpha}
)^{2}
} - 
\frac{2
(
{\rm e}^{\mu + \frac{\sigma^{2}}{2}}
\int_{\frac{\mu + \sigma^{2}-\log m_{\times}}{\sigma}}^{\infty}
Dx H(x+ \sigma)
-
\frac{(\alpha-1)^{2}m_{\times}^{3-2\alpha}}
{(\alpha-2)(2\alpha-3)}
)}
{
(
H(\frac{\mu -\log m_{\times}}{\sigma}) + 
m_{\times}^{1-\alpha}
)
(
{\rm e}^{\mu + \frac{\sigma^{2}}{2}}
H
(\frac{\mu + \sigma^{2}-\log m_{\times}}{\sigma})
+
(
\frac{\alpha-1}{\alpha-2}
)
m_{\times}^{2-\alpha}
)
}  
\end{equation}
and $k$ index is given as a solution of 
\begin{eqnarray} 
H^{-1}
\left[
\left\{
H\left(
\frac{\mu + \sigma^{2}-\log m_{\times}}{\sigma}
\right)
+
\left(
\frac{\alpha-1}{\alpha-2}
\right)
{\rm e}^{-\mu -\frac{\sigma^{2}}{2}}
m_{\times}^{1-\alpha}
\right\} (1-k)
\right]  & - & 
H^{-1}
\left[
\left\{
H
\left(
\frac{\mu -\log m_{\times}}{\sigma}
\right)
+m_{\times}^{1-\alpha}
\right\}k
\right] \nonumber \\
\mbox{} & = & \sigma, \quad 
0 \leq X \leq 
\frac{H(\frac{\mu -\log m_{\times}}{\sigma})}
{H(\frac{\mu-\log m_{\times}}{\sigma})+
m_{\times}^{1-\alpha}},
\end{eqnarray}
and, 
\begin{equation}
k =  
\frac{
(
\frac{\alpha-1}{\alpha-2}
)
(H
(
\frac{\mu-\log m_{\times}}{\sigma}
) + m_{\times}^{1-\alpha}
)^{\frac{2-\alpha}{1-\alpha}}
}
{
{\rm e}^{\mu + \frac{\sigma^{2}}{2}}
H
(
\frac{\mu + \sigma^{2}-\log m_{\times}}{\sigma}
)+
(
\frac{\alpha-1}{\alpha-2}
)
m_{\times}^{2-\alpha}
} 
(1-k)^{\frac{2-\alpha}{1-\alpha}}, \quad \quad 
\frac{H(\frac{\mu -\log m_{\times}}{\sigma})}
{H(\frac{\mu-\log m_{\times}}{\sigma})+
m_{\times}^{1-\alpha}}
< X \leq 1.
\end{equation}
\subsection{(c) Uniform distribution follows a power law distribution}
We next consider the case 
$F_{1}(m)=1/a$ and 
$F_{2}(m)=(\alpha-1)m^{-\alpha}$ with 
$m_{0}=0$. 
For this case, we have 
\begin{eqnarray}
Q_{1}(r ) & = & 
\frac{r}{a},\quad 
Q_{2}(r )=r^{1-\alpha}; \\
R_{1}(r ) & = & 
\frac{r^{2}}{2a},\quad 
R_{2}(r )= 
\frac{(\alpha-2)r^{2-\alpha}}{(\alpha-2)}, 
\end{eqnarray}
and using the staffs 
$Q_{1}(m_{\times})=m_{\times}/a, 
Q_{1}(m_{0})=Q_{2}(\infty)=0, 
Q_{2}(m_{\times})=m_{\times}^{1-\alpha}$ and 
$R_{1}(m_{\times})=m_{\times}^{2}/2a, 
R_{2}(m_{\times})=(\alpha-1)m_{\times}/(\alpha-2)$, 
and accompanying 
\begin{eqnarray}
S_{1}(m_{0},m_{\times}) & = & 
\frac{m_{\times}^{3}}{6a^{2}}, \\
T_{2}(m_{\times}) & = & -\frac{(\alpha-1)^{2}m_{\times}^{3-2\alpha}}
{(\alpha-2)(\alpha-3)}, 
\end{eqnarray}
the Lorenz curve is explicitly given as 
\begin{equation}
Y = 
\left\{
\begin{array}{lc}
\frac{(m_{\times}+am_{\times}^{1-\alpha})^{2}}
{
m_{\times}^{2}+
\frac{2a(\alpha-1)}{(\alpha-2)}
m_{\times}^{2-\alpha}}
X^{2},
 & 
\quad \quad 
0 \leq X \leq 
\frac{m_{\times}}{m_{\times}+am_{\times}^{1-\alpha}},
\\
1- 
\frac{
\frac{2a^{\frac{1}{\alpha-1}}(\alpha-1)}{(\alpha-2)}
(
m_{\times}+
am_{\times}^{1-\alpha}
)^{\frac{2-\alpha}{1-\alpha}}
}
{m_{\times}^{2}
+
\frac{2a(\alpha-1)}{(\alpha-2)}
m_{\times}^{2-\alpha}
} 
(1-X)^{\frac{2-\alpha}{1-\alpha}},
& 
\quad \quad 
\frac{m_{\times}}{m_{\times}+am_{\times}^{1-\alpha}}
< X \leq 1.
\end{array}
\right.
\end{equation}
The $g$ index is calculated as 
\begin{equation}
g = 
\frac{m_{\times}^{2}-am_{\times}^{2-2\alpha}}
{
(
m_{\times} + am_{\times}^{1-\alpha})^{2}
}
-
\frac{
2
(
m_{\times}^{3}
-
\frac{6a^{2}(\alpha-1)^{2}m_{\times}^{3-2\alpha}}{(\alpha-2)(2\alpha-3)}
)
}
{3 (m_{\times} +am_{\times}^{1-\alpha})
(
m_{\times}^{2}
+
\frac{2a(\alpha-1)}{(\alpha-2)}
m_{\times}^{2-\alpha}
)
},
\end{equation}
and $k$ index is determined by the solution of 
\begin{equation}
k = 
\left\{
\begin{array}{lc}
\frac{
\left(
m_{\times}^{2}+
\frac{2a(\alpha-1)}{(\alpha-2)}
m_{\times}^{2-\alpha}
\right)
\left(
\sqrt{
1+
\frac{4(m_{\times}+am_{\times}^{1-\alpha})^{2}}
{m_{\times}^{2}+
\frac{2a(\alpha-1)}{(\alpha-2)}
m_{\times}^{2-\alpha}}
}-1
\right)
}
{2(m_{\times}+am_{\times}^{1-\alpha})^{2}}, & 
\quad \quad 
0 \leq k \leq 
\frac{m_{\times}}{m_{\times}+am_{\times}^{1-\alpha}},
\\
\frac{
\frac{2a^{\frac{1}{\alpha-1}}(\alpha-1)}
{(\alpha-2)}
(
m_{\times}
+
am_{\times}^{1-\alpha}
)^{\frac{2-\alpha}{1-\alpha}}
}
{
m_{\times}^{2}
+
\frac{2a (\alpha-1)}{(\alpha-2)}
m_{\times}^{2-\alpha}
} 
(1-k)^{\frac{2-\alpha}{1-\alpha}}, & 
\quad \quad 
\frac{m_{\times}}{m_{\times}+am_{\times}^{1-\alpha}} 
< k \leq 1.
\end{array}
\right. 
\end{equation}
\subsection{(d) Uniform distribution with a lognormal tail}
Here we choose 
$F_{1}(m)=1/a, F_{2}(m)=
{\rm e}^{-\frac{(\log m-\mu)^{2}}{2\sigma^{2}}}/\sqrt{2\pi}\sigma m$ with 
$m_{0}=0$. 
We have 
\begin{eqnarray}
Q_{1} (r ) & = & 
\frac{r}{a},\,\,\,
Q_{2}( r) = 
H
\left(
\frac{\log r-\mu}{\sigma}
\right); \\
R_{1}(r ) & = & 
\frac{r^{2}}{2a},\,\,\,
R_{2}(r ) = 
{\rm e}^{\mu + \frac{\sigma^{2}}{2}}
H
\left(
\frac{\log r -\mu -\sigma^{2}}{\sigma}
\right),
\end{eqnarray}
and by making use of the staffs 
$Q_{1}(m_{\times})=m_{\times}/a, 
Q_{1}(m_{0})=Q_{2}(\infty)=0, 
Q_{2}(m_{\times})=H(\frac{\log m_{\times}-\mu}{\sigma})$ and 
$R_{1}(m_{\times})=m_{\times}^{2}/2a, 
R_{2}(m_{\times})={\rm e}^{\mu+ \frac{\sigma^{2}}{2}}
H(\frac{\log m_{\times}-\mu-\sigma^{2}}{\sigma})$, 
and accompanying 
\begin{eqnarray}
S_{1}(m_{0},m_{\times}) & = & 
\frac{m_{\times}^{3}}{6a^{2}},  \\
T_{2}(m_{\times}) & = & 
-{\rm e}^{\mu + \frac{\sigma^{2}}{2}}
\int_{\frac{\log m_{\times}-\mu}{\sigma}}^{\infty}
Dx H(x-\sigma), 
\end{eqnarray}
the Lorenz curve is given by 
\begin{equation}
Y = 
\left\{
\begin{array}{lc}
\frac{(m_{\times} + 
aH
(\frac{\log m_{\times} -\mu}{\sigma})
)^{2}}
{
m_{\times}^{2}+
2a {\rm e}^{\mu + \frac{\sigma^{2}}{2}}
H(\frac{\log m_{\times}-\mu -\sigma^{2}}{\sigma})
} X^{2}, & 
\quad \quad 
0 \leq X \leq 
\frac{m_{\times}}
{m_{\times}+
aH
(\frac{\log m_{\times}-\mu}{\sigma})
},
\\
1- 
\frac{2a 
{\rm e}^{\mu + 
\frac{\sigma^{2}}{2}}
H(
H^{-1}[
\frac{1}{a} \{
m_{\times}
+aH(\frac{\log m_{\times}-\mu}{\sigma})
\}
(1-X)]-\sigma)
}
{m_{\times}^{2}+
2a {\rm e}^{\mu + \frac{\sigma^{2}}{2}}
H(\frac{\log m_{\times}-\mu-\sigma^{2}}{\sigma})}, & 
\quad \quad 
\frac{m_{\times}}
{m_{\times}+
aH
(\frac{\log m_{\times}-\mu}{\sigma})
}
<X \leq 1.
\end{array}
\right. 
\end{equation}
The $g$ index is obtained as 
\begin{equation}
g = 
\frac{m_{\times}^{2}-\{2aH(\frac{\log m_{\times}-\mu}{\sigma})\}^{2}}
{(m_{\times}+2aH (\frac{\log m_{\times}-\mu}{\sigma})^{2}}-
\frac{
2(m_{\times}^{3}
-6a^{2}{\rm e}^{\mu + \frac{\sigma^{2}}{2}}
\int_{\frac{\log m_{\times}-\mu}{\sigma}}^{\infty}
Dx H(x-\sigma))
}
{
3(m_{\times}+2aH(\frac{\log m_{\times}-\mu}{\sigma}))
(m_{\times}^{2}+
2a {\rm e}^{\mu + \frac{\sigma^{2}}{2}}
H(\frac{\log m_{\times}-\mu -\sigma^{2}}{\sigma}))},
\end{equation}
and $k$ index becomes
\begin{equation}
k = 
\frac{
(m_{\times}^{2}+
2a {\rm e}^{\mu + \frac{\sigma^{2}}{2}}
H(\frac{\log m_{\times}-\mu -\sigma^{2}}{\sigma}))
\sqrt{
1+
\frac{4(m_{\times}+aH(\frac{\log m_{\times}-\mu}{\sigma}))^{2}}
{(m_{\times}^{2}+
2a {\rm e}^{\mu + \frac{\sigma^{2}}{2}}
H(\frac{\log m_{\times}-\mu -\sigma^{2}}{\sigma}))}
-1}
}
{2(m_{\times}+aH(\frac{\log m_{\times}-\mu}{\sigma}))^{2}}, \quad \quad 
0 \leq k \leq 
\frac{m_{\times}}
{m_{\times}+aH(\frac{\log m_{\times}-\mu}{\sigma})}.
\end{equation}
and
\begin{eqnarray}
k= H^{-1}\left[
\frac{1}{a}
\left\{
m_{\times}+aH
\left(
\frac{\log m_{\times}-\mu}{\sigma}
\right)
\right\}(1-k)
\right] & - & 
H^{-1}
\left[
\frac{1}{2a{\rm e}^{\mu+\frac{\sigma^{2}}{2}}}
\left\{
m_{\times}^{2}+
2a {\rm e}^{\mu + \frac{\sigma^{2}}{2}}
H
\left(
\frac{\log m_{\times}-\mu -\sigma^{2}}{\sigma}
\right)
\right\}k
\right] \nonumber \\
\mbox{} & = & \sigma, \quad \quad 
\frac{m_{\times}}
{m_{\times}+aH(\frac{\log m_{\times}-\mu}{\sigma})}
<k \leq 1.
\end{eqnarray}
\subsection{(e) Double power laws}
We consider the case 
$F_{1}(m)=(\alpha-1)m^{-\alpha}$ and 
$F_{2}(m)=(\beta-1)m^{-\beta}$ with 
$m_{0}=1$. Then,
we have 
\begin{eqnarray}
Q_{1}(r ) & = & 1-r^{1-\alpha},\quad Q_{2}(r )  =  
r^{1-\beta}; \\
R_{1}(r ) & = & 
\left(
\frac{\alpha-1}{\alpha-2}
\right)(1-r^{2-\alpha}),\quad 
R_{2}(r ) = 
\left(
\frac{\beta-1}{\beta-2}
\right)r^{2-\beta}.
\end{eqnarray}
Using the staffs 
$Q_{1}(m_{\times})=1-m_{\times}^{1-\alpha},
Q_{1}(m_{0})=Q_{2}(\infty)=0, 
Q_{2}(m_{\times})=m_{\times}^{1-\beta}$ and 
$R_{1}(m_{\times})=(\alpha-1)(1-m_{\times}^{2-\alpha})/(\alpha-2), 
R_{2}(m_{\times})=(\beta-1) m_{\times}^{2-\beta}/(\beta-2)$, 
accompanying 
\begin{eqnarray}
S_{1}(m_{0},m_{\times}) & = & 
\frac{\alpha-1}{2\alpha-3}
-
\frac{(\alpha-1)(2\alpha-3)m_{\times}^{1-\alpha}
-(\alpha-1)^{2}m_{\times}^{3-2\alpha}}
{(\alpha-2)(2\alpha-3)},  \\
T_{2}(m_{\times}) & = & 
-\frac{(\beta-1)^{2}m_{\times}^{3-2\beta}}
{(\beta-2)(2\beta-3)}, 
\end{eqnarray}
 the Lorenz curve is obtained as 
\begin{equation}
Y = 
\left\{
\begin{array}{lc}
\frac{1-
[
1-(1-m_{\times}^{1-\alpha}
+m_{\times}^{1-\beta})X
]^{\frac{2-\alpha}{1-\alpha}}
}
{
1-m_{\times}^{2-\alpha}+
\frac{(\alpha-2)(\beta-1)}
{(\alpha-1)(\beta-2)}
m_{\times}^{2-\beta}
},  & 
\quad \quad 
0 \leq X \leq 
\frac{1-m_{\times}^{1-\alpha}}
{1-m_{\times}^{1-\alpha}+m_{\times}^{1-\beta}},
\\
1-
\frac{
\frac{(\alpha-2)(\beta-1)}
{(\alpha-1)(\beta-2)}
[
(1-m_{\times}^{1-\alpha}
+m_{\times}^{1-\beta})(1-X)]^{\frac{2-\beta}{1-\beta}}
}
{1-m_{\times}^{2-\alpha}
+
\frac{(\alpha-2)(\beta-1)}
{(\alpha-1)(\beta-2)}
m_{\times}^{2-\beta}
}, & 
\quad \quad 
\frac{1-m_{\times}^{1-\alpha}}
{1-m_{\times}^{1-\alpha}+m_{\times}^{1-\beta}} <X \leq 1.
\end{array}
\right. 
\end{equation}
The $g$ index is given  by 
\begin{equation}
g = 
\frac{(1-m_{\times}^{1-\alpha})^{2}-m_{\times}^{2-2\beta}}
{
(1-m_{\times}^{1-\alpha}+
m_{\times}^{1-\beta})^{2}}
-
\frac{2(
\frac{\alpha-2}{\alpha-1})
\{
\frac{\alpha-1}{2\alpha-3}
-
\frac{(\alpha-1)(2\alpha-3)m_{\times}^{1-\alpha}
-(\alpha-1)^{2}m_{\times}^{3-2\alpha}}
{(\alpha-2)(2\alpha-3)}
-
\frac{(\beta-1)^{2}m_{\times}^{3-2\beta}}
{(\beta-2)(2\beta-3)}
\}
}
{
(1-m_{\times}^{1-\alpha}+m_{\times}^{1-\beta})
(1-m_{\times}^{2-\alpha}
+
\frac{(\alpha-2)(\beta-1)}
{(\alpha-1)(\beta-2)}
m_{\times}^{2-\beta})
}, 
\end{equation}
and the $k$ index as 
\begin{equation}
k = 
\left\{
\begin{array}{cl}
\frac{
1-[1
-\{1-m_{\times}^{2-\alpha}
+
\frac{(\alpha-2)(\beta-1)}
{(\alpha-1)(\beta-2)}
m_{\times}^{2-\beta}
\}(1-k)
]^{\frac{1-\alpha}{2-\alpha}}
}
{
1-m_{\times}^{1-\alpha}+m_{\times}^{1-\beta}
}, & 
\quad \quad 
0 \leq X \leq 
\frac{1-m_{\times}^{1-\alpha}}
{1-m_{\times}^{1-\alpha}+m_{\times}^{1-\beta}},
\\
\frac{
\frac{(\alpha-2)(\beta-1)}
{(\alpha-1)(\beta-2)}
[
\{
1-m_{\times}^{1-\alpha}+
m_{\times}^{1-\beta}
\}(1-k)]^{\frac{2-\beta}{1-\beta}}
}
{1-m_{\times}^{2-\alpha}+m_{\times}^{2-\beta}}, & 
\quad \quad 
\frac{1-m_{\times}^{1-\alpha}}
{1-m_{\times}^{1-\alpha}+m_{\times}^{1-\beta}} <X \leq 1.
\end{array}
\right.
\end{equation}
\subsection{(f) Power law distribution with a lognormal tail}
Finally we consider the case 
$F_{1}(m)=(\alpha-1)m^{-\alpha}$ and 
$F_{2}(m)=
{\rm e}^{-\frac{(\log m-\mu)^{2}}{2\sigma^{2}}}/\sqrt{2\pi}\sigma m$ with 
$m_{0}=1$. 
Then, we have 
\begin{eqnarray}
Q_{1}(r ) & = & 
1-r^{1-\alpha},\quad 
Q_{2}(r ) = 
H
\left(
\frac{\log r-\mu}{\sigma}
\right); \\
R_{1}(r ) & = & 
\left(
\frac{\alpha-1}{\alpha-2}
\right)(1-r^{2-\alpha}),\quad 
R_{2}(r ) = 
{\rm e}^{\mu + \frac{\sigma^{2}}{2}}
H
\left(
\frac{\log r -\mu -\sigma^{2}}{\sigma}
\right).
\end{eqnarray}
Using the staffs 
$Q_{1}(m_{\times})=1-m_{\times}^{1-\alpha},
Q_{1}(m_{0})=Q_{2}(\infty)=0,
Q_{2}(m_{\times})=H(\frac{\log m_{\times}-\mu}{\sigma})$ and 
$R_{1}(m_{\times})=(\alpha-1)(1-m_{\times}^{2-\alpha})/(\alpha-2), 
R_{2}(m_{\times})={\rm e}^{\mu + \frac{\sigma^{2}}{2}}
H(\frac{\log m_{\times}-\mu-\sigma^{2}}{\sigma})$, 
accompanying 
\begin{eqnarray}
S_{1}(m_{0},m_{\times}) & = & 
\frac{\alpha-1}{2\alpha-3}
-
\frac{(\alpha-1)(2\alpha-3)m_{\times}^{1-\alpha}
-(\alpha-1)^{2}m_{\times}^{3-2\alpha}}
{(\alpha-2)(2\alpha-3)},  \\
T_{2}(m_{\times}) & = & 
-{\rm e}^{\mu + \frac{\sigma^{2}}{2}}
\int_{\frac{\log m_{\times}-\mu}{\sigma}}^{\infty}
Dx H(x-\sigma), 
\end{eqnarray}
the Lorenz curve is given by 
\begin{equation}
Y = 
\left\{
\begin{array}{lc}
\frac{1-[
1-\{
1-m_{\times}^{1-\alpha}
+H
(\frac{\log m_{\times}-\mu}{\sigma})
\}X]^{\frac{2-\alpha}{1-\alpha}}}
{1-m_{\times}^{2-\alpha}
+
{\rm e}^{\mu + \frac{\sigma^{2}}{2}}
(
\frac{\alpha-2}{\alpha-1}
)H
(\frac{\log m_{\times}-\mu-\sigma^{2}}{\sigma})}, & 
\quad \quad 
0 \leq X \leq 
\frac{1-m_{\times}^{1-\alpha}}
{1-m_{\times}^{1-\alpha}+
H(\frac{\log m_{\times}-\mu -\sigma^{2}}{\sigma})
},
\\
1-
\frac{
(\frac{\alpha-2}{\alpha-1})
{\rm e}^{\mu + \frac{\sigma^{2}}{2}}
H(
H^{-1}[
\{
1-m_{\times}^{1-\alpha}
+
H(\frac{\log m_{\times}-\mu}{\sigma})
\}(1-X)]-\sigma)
}
{
1-m_{\times}^{2-\alpha}
+
{\rm e}^{\mu + \frac{\sigma^{2}}{2}}
(\frac{\alpha-2}{\alpha-1})
H(
\frac{\log m_{\times}-\mu -\sigma^{2}}{\sigma}
)}, & 
\quad \quad 
\frac{1-m_{\times}^{1-\alpha}}
{1-m_{\times}^{1-\alpha}+
H(\frac{\log m_{\times}-\mu -\sigma^{2}}{\sigma})
} < X \leq 1.
\end{array}
\right.
\end{equation}
and the $g$ index is given by 
\begin{equation}
g = 
\frac{(1-m_{\times}^{1-\alpha})^{2}-
H(\frac{\log m_{\times}-\mu}{\sigma})^{2}}
{
(1-m_{\times}^{1-\alpha}+
H(\frac{\log m_{\times}-\mu}{\sigma})
)^{2}
}-
\frac{2(\frac{\alpha-2}{\alpha-1})
\{
\frac{\alpha-1}{2\alpha-3}
-
\frac{(\alpha-1)(2\alpha-3)m_{\times}^{1-\alpha}
-(\alpha-1)^{2}m_{\times}^{3-2\alpha}}
{(\alpha-2)(2\alpha-3)}
-{\rm e}^{\mu + \frac{\sigma^{2}}{2}}
\int_{\frac{\log m_{\times}-\mu}{\sigma}}^{\infty}
Dx H(x-\sigma)
\}
}
{
(1-m_{\times}^{1-\alpha}
+H(
\frac{\log m_{\times}-\mu}{\sigma}
))
(
1-m_{\times}^{2-\alpha}
+
{\rm e}^{\mu + \frac{\sigma^{2}}{2}}
(\frac{\alpha-2}{\alpha-1})
H(\frac{\log m_{\times}-\mu -\sigma^{2}}{\sigma})
)
}, 
\end{equation}
and $k$ index is  
\begin{equation}
k = 
\frac{1-[
1-\{
1-m_{\times}^{2-\alpha}+
{\rm e}^{\mu + \frac{\sigma^{2}}{2}}
(\frac{\alpha-2}{\alpha-1})
H(\frac{\log m_{\times}-\mu -\sigma^{2}}{\sigma})
\}(1-k)]^{\frac{1-\alpha}{2-\alpha}}}
{
1-m_{\times}^{1-\alpha}+
H(\frac{\log m_{\times}-\mu}{\sigma})
}, \quad \quad 
0 \leq k \leq 
\frac{1-m_{\times}^{1-\alpha}}
{1-m_{\times}^{1-\alpha}
+
H(\frac{\log m_{\times}-\mu}{\sigma})}.
\end{equation}
and
\begin{eqnarray}
\mbox{} &&  k= H^{-1}
\left[
\left\{
1-m_{\times}^{1-\alpha}
+
H 
\left(
\frac{\log m_{\times}-\mu}{\sigma}
\right)
\right\}(1-k)
\right] \nonumber \\
\mbox{} & - & 
H^{-1}
\left[
\frac{1}{{\rm e}^{\mu + \frac{\sigma^{2}}{2}}
(\frac{\alpha-2}{\alpha-1})}
\left\{
1-m_{\times}^{2-\alpha}
+
{\rm e}^{\mu + \frac{\sigma^{2}}{2}}
\left(\frac{\alpha-2}{\alpha-1}\right)
H\left(\frac{\log m_{\times}-\mu -\sigma^{2}}{\sigma}\right)
\right\}k
\right], \nonumber \\
\mbox{} & \mbox{} & 
\frac{1-m_{\times}^{1-\alpha}}
{1-m_{\times}^{1-\alpha}
+
H(\frac{\log m_{\times}-\mu}{\sigma})} < k \leq 1.
\end{eqnarray}

\end{document}